\documentclass[twocolumn,showpacs,amsmath,amssymb,floatfix,prb,showkeys]{revtex4}

\usepackage{graphicx}
\usepackage{epsfig} 
\usepackage{dcolumn}
\usepackage{bm}
\usepackage{times}

\begin{document} 


\title{Disorder-quenched Kondo effect in mesosocopic electronic
systems}

\author{Stefan Kettemann$^{1}$} 

\author{Eduardo R. Mucciolo$^2$} 

\affiliation{$^1$ Institut f\"ur Theoretische Physik, Universit\" at
Hamburg, Jungiusstra\ss e 9, 20355 Hamburg, Germany,}

\affiliation{$^2$ Department of Physics, University of Central
Florida, P.O. Box 162385, Orlando, FL 32816-2385, USA}

\date{\today}

\begin{abstract}
Nonmagnetic disorder is shown to quench the screening of magnetic
moments in metals, the Kondo effect. The probability that a magnetic
moment remains free down to zero temperature is found to increase with
disorder strength. Experimental consequences for disordered metals are
studied. In particular, it is shown that the presence of magnetic
impurities with a small Kondo temperature enhances the electron's
dephasing rate at low temperatures in comparison to the clean metal
case. It is furthermore proven that the width of the distribution of
Kondo temperatures remains finite in the thermodynamic (infinite
volume) limit due to wave function correlations within an energy
interval of order $1/\tau$, where $\tau$ is the elastic scattering
time. When time-reversal symmetry is broken either by applying a
magnetic field or by increasing the concentration of magnetic
impurities, the distribution of Kondo temperatures becomes narrower.
\end{abstract}



\maketitle

\section{Introduction} 
\label{sec:introduction}

A local magnetic impurity changes the ground state of a Fermi liquid
due to the correlations created by the antiferromagnetic exchange
interaction between its localized spin and the delocalized
electrons.\cite{hewson,abrikosov} At temperatures below the Kondo
temperature $T_K$ the spin of the magnetic impurity is screened by the
formation of a singlet state with the conduction
electrons.\cite{nozieres,wilson} Disorder affects the formation of
this Kondo singlet in various ways.\cite{mirandareview} In particular,
the Kondo temperature may depend on the positioning of the magnetic
impurities in the host lattice. Thus, for a sample containing many
magnetic impurities, $T_K$ may be distributed due to fluctuations of
the microscopic exchange coupling.\cite{bhatt,langenfeld,kbu03}
Impurities (magnetic or nonmagnetic) and lattice defects also cause
fluctuations in the local density of states (LDOS) of the conduction
electrons at the magnetic impurity site. Thus, the distributions of
the Kondo temperature and of the LDOS are related in a disordered
metal.\cite{miranda,jetp}

In this paper we explore the consequences of nonmagnetic disorder for
the Kondo effect. We base our study on the statistical properties of
$T_K$ for the different dynamical regimes of a disordered metal. Note
that the Kondo temperature accounts for a crossover rather than a
sharp transition in the metal properties and that may raise the
question whether $T_K$ is sufficiently well defined. In weakly
disordered metals it is possible in principle to extract $T_K$ by
fitting physical quantities such as the spin susceptibility and the
electron dephasing rate versus temperature against the corresponding
universal scaling function. While that scaling function may be
modified itself by the disorder,\cite{kbu03,future} the Kondo
screening is still governed by $T_K$, and one may study its
sample-to-sample fluctuations. Below, we show that these fluctuations
may have a significant impact on the low-temperature properties of a
metallic sample even in the thermodynamic, infinite volume limit.

\begin{figure}
\includegraphics[width=6cm]{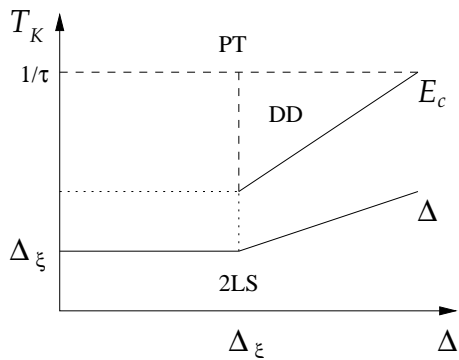}
\caption{Overview of the different regimes of a mesoscopic Kondo
system of linear size $L$, as parametrized by the system's mean level
spacing $\Delta=1/\nu L^d$ and the unaveraged Kondo temperature of a
single magnetic impurity $T_K$. $E_c$ is the Thouless energy, $1/\tau$
is the elastic scattering rate, and $\Delta_\xi = 1/\nu \xi^d$, with
$\xi$ being the localization length. The acronyms denote: Two levels
coupled to the magnetic impurity (2LS), diagrammatic expansion in
diffusion ladder diagrams (DD), and perturbative regime (PT), with
$1/g= (E_F \tau)^{-1}$ as the small expansion parameter. We only
consider metallic systems where the relation $\Delta_\xi \ll 1/\tau
\ll E_F$ is fulfilled.}
\label{fig:kondoboxmap}
\end{figure}

It is useful to compare the magnitude of $T_K$ to other relevant
energy scales of a metallic sample. These scales are the mean level
spacing $\Delta = 1/\nu\, L^d$, the Thouless energy $E_c=D_e/L^2$, and
the elastic scattering rate $1/\tau$. Here, $\nu$ denotes the density
of states at the Fermi level, $D_e =v_F^2 \tau/d$ is the diffusion
constant, and $d$ is the dimensionality of diffusion (we have set
$\hbar=k_B=1$). This comparison allows one to establish several
distinct regimes, as sketched in Fig. \ref{fig:kondoboxmap}. We note
that, in practice, $T_K$ does not exceed $1/\tau$ in metals, therefore
standard perturbation theory in the disorder potential may not be used
to describe the effect of disorder on the Kondo effect. However, there
is a large regime of experimental interest where a diagrammatic
expansion may be used, namely, $E_c < T_K < 1/\tau$.\cite{shklovskii}
For small samples, when $\Delta < T_K < E_c$, random matrix theory
(RMT) can be applied and the distribution of Kondo temperatures is
expected to scale with $\Delta$ alone. In the opposite limit, when the
sample is so large that the localization length $\xi$ is smaller than
the linear size $L$, the Thouless energy and the global level spacing
are irrelevant and the Kondo temperature is determined by the mean
level spacing for states localized in the vicinity of the magnetic
impurity, $\Delta_\xi = 1/\nu\, \xi^d$.\cite{alk95,altland,sk99} In
this case, one can expect the distribution of Kondo temperatures to
scale with $\Delta_\xi$ instead of $\Delta$. Finally, when $T_K$ is
smaller than the spacing between neighboring energy levels at the
Fermi surface (either $\Delta$ or $\Delta_\xi$), the distribution of
$T_K$ will be mainly determined by the fluctuations of the wave
functions and level spacing of the two eigenstates closest to the
Fermi energy. Accordingly, in this regime the Kondo temperature is
determined by the coupling of the magnetic impurity to a two-level
system (2LS).

Our analytical calculations show that the average Kondo temperature is
enhanced by weak disorder. Disorder also induces a {\it finite width}
in the distribution of $T_K$ which is shown to persist even in the
infinite sample, thermodynamic limit. This effect is due to the
existence of local correlations between eigenfunctions at different
energies within a macroscopically large energy interval
$1/\tau$. These spectral correlations also lead to a bimodal structure in
the distribution of $T_K$, with a long tail stretching toward
$T_K=0$. The numerical simulations show that for a sizeable number of
disorder configurations, no finite value of $T_K$ exist within the
lowest order self-consistent perturbation theory, indicating that in
those cases the magnetic moment is not screened by the conduction
electrons. We find that there is a finite probability of having free
moments even for values of $J$ such that the average Kondo temperature
$\langle T_K \rangle $ is much larger than the mean level spacing
$\Delta$.

For small metallic grains, it is well-know that RMT provides a very
good description of the statistical fluctuations of single-particle
states. We therefore use RMT to investigate analytically and
numerically the statistics of $T_K$ in such  systems. The ratio
between the standard deviations of $T_K$ for the orthogonal and
unitary ensemble is found numerically to be equal to $\sqrt{2}$ for a
wide range of exchange couplings $J$, in agreement with analytical
calculations. Furthermore, it is shown numerically that a finite
number of free moments persists to exist in this regime. This result
is consistent with an analytical calculation based on the 2LS
approximation.

In regard to the width of the distribution of $T_K$ in the RMT regime,
we find a counter-intuitive result at first glance, namely, that the
distribution becomes {\it narrower} when increasing the
concentration. This effect maybe understood as a consequence of the
suppression of the weak localization correction by the breaking of
time-reversal invariance, which in RMT is seen as a reduction in wave
function fluctuations. We also consider whether additional
correlations between wave functions induced by breaking of
time-reversal symmetry\cite{adam} within an interval of order of the
spin scattering rate $1/\tau_s$ can lead to an enhancement of the
width of the distribution of $T_K$. We find that while the width
increases by a factor $\sqrt{T_K \tau_s}$ as compared to the pure RMT
ensembles, it nevertheless vanishes in the thermodynamic limit when
$\Delta \rightarrow 0$, and therefore is negligible in comparison with
the magnitude of the broadening caused by wave functions correlations
beyond RMT.

The large fluctuations of $T_K$ in disordered metals impact the
behavior of several thermodynamic and transport properties at low
temperatures. We consider in particular the influence of these
fluctuations on the temperature dependence of the electron's dephasing
rate. We find that nonmagnetic disorder combined with spin scattering
due to magnetic impurities increases the dephasing rate at
temperatures $T< \langle T_K \rangle$. The dephasing rate is readily
accessible in transport experiments by measuring weak-localization
corrections to the resistance. Our predictions are in qualitative
agreement with recent experimental results.

Our work details and extends the discussions provided in a previous
letter (Ref. \onlinecite{jetp}) and also includes new numerical and
analytical results. The presentation is organized in the following
way. In Sec. \ref{sec:formulation} we introduce the problem of finding
the distribution of the Kondo temperature as defined by an integral
equation. The central analytical and numerical results of our paper are
presented in Secs. \ref{sec:disorderedmetals} and \ref{sec:numerics},
respectively. There, the distribution, average, and standard deviation
of the Kondo temperature are studied as function of disorder strength
by taking eigenstates and eigenenergies from a disordered
tight-binding model. The average and standard deviation are derived
analytically as function of $\Delta, E_c$, and the disorder strength
and compared with the numerical results. We also study how the number
of free moments depends on the exchange coupling and disorder
strength.

In Sec. \ref{sec:rmt}, we consider the RMT regime. We present the
analytical results obtained for the distribution of $T_K$ by taking
into account only wave function fluctuations. We compare analytical
estimates of the average and standard deviation of $T_K$ with
numerical results obtained from simulations of random matrix
ensembles. Both large- and small-$T_K$ limit are considered.

The dependence of the distribution of the Kondo temperature on the
concentration of magnetic impurities is studied in
Sec. \ref{sec:concentration}. In Sec. \ref{sec:detection} we describe
how the fluctuations of $T_K$ affect the dephasing time of electrons
in metallic thin films and quasi-one-dimensional wires. Finally, in
Sec. \ref{sec:conclusions} we summarize our results and make our
concluding remarks. Appendix A includes the derivation of the integral
equation for the Kondo temperature in a one-loop approximation for a
disordered sample, together with a critical discussion of its
validity. In Appendix B, for reference, we briefly present the
solution of the integral equation in the clean limit using a
picket-fence spectrum (equidistant energy levels).

\section{Formulation of the problem}
\label{sec:formulation}

We will consider a single magnetic impurity in an isolated,
phase-coherent metallic sample where the energy levels are well
separated, a system previously referred in the literature as the Kondo
box.\cite{kroha,suga} To determine $T_K$, we will adopt the following
self-consistent equation,\cite{nagaoka,fulde,jetp}
\begin{equation}
\label{eq:tk} 
1 = \frac{J\, }{2 N} \sum_{n=1}^N \frac{\Omega |\psi_n(0)|^2} {E_n -
  E_F} \tanh \left( \frac{E_n-E_F}{2 T_K} \right),
\end{equation} 
where $N$ is the number of states in the spectrum, $\Omega=L^d$ is the
sample volume, $E_F$ is the Fermi energy, and $J$ is the exchange
coupling between the magnetic impurity and the delocalized
electrons. The eigenenergies and eigenfunctions of the sample are,
respectively, $E_n$ and $\psi_n({\bf r})$, while the impurity is
located at ${\bf r}=0$. Equation (\ref{eq:tk}) is obtained from
second-order perturbation theory in Appendix \ref{sec:appendixA}. A
similar expression can be obtained from the zero-temperature,
self-consistent solution of the one-loop renormalization group (RG)
equation.\cite{anderson,wilson} However, in that case one misses the
$\tanh$ factor (which accounts for the finite-temperature occupation
numbers) and the result is only valid for $T_K \gg \Delta$. While the
approximations involved in deriving Eq. (\ref{eq:tk}) are not
sufficient for describing the properties of the system below $T_K$, it
is important to remark that Eq. (\ref{eq:tk}) does yield a good
estimate for the Kondo temperature, which is the relevant scale for
the low-temperature behavior of Kondo systems. The two-loop correction
has been found to change the Kondo temperature by a factor
$\sqrt{J/D}$.\cite{wilson} In the thermodynamic limit (infinite
volume), the physics at temperatures $T \ll T_K$ is known to be that
of an effective Fermi liquid, where the Kondo temperature determines
the Landau parameters.\cite{nozieres} For example, the effective mass,
the density of states, and thereby the specific heat become enhanced
by a factor $1+n_m/(\pi \nu T_K)$, while the paramagnetic
susceptibility is enhanced by a factor $1+2 n_m/(\pi \nu T_K)$, where
$n_m$ is the concentration of magnetic
impurities.\cite{nozieres,abrikosov}

It is convenient to rescale Eq. (\ref{eq:tk}) to a dimensionless form,
in which case it becomes
\begin{equation} 
\label{eq:sctk}
1 = \frac{1}{2x} \sum_{n=1}^N \frac{x_n}{s_n} \tanh \left(
\frac{s_n}{2 \kappa} \right),
\end{equation} 
where $x=D/J$, $\kappa = T_{\rm K}/D$, $x_n = \Omega |\psi_n ({\bf r})
|^2$ is the probability density of the $n$th eigenstate, and $s_n =
(E_n-E_F)/\Delta$ is the corresponding eigenenergy measured relative
to the Fermi energy in units of $\Delta$. In the following, we will
assume that the Fermi energy is in the middle of two energy levels, so
that the number of electrons in the Kondo box is even. (For an odd
number of electrons, the unpaired electron at the Fermi energy forms a
singlet with the magnetic impurity with binding energy $J$ and the
problem becomes trivial.) Using Eq. (\ref{eq:sctk}), the distribution
of Kondo temperatures is determined by
\begin{equation} 
\label{eq:ptk}
P(T_K) = \left| \frac{d}{d T_K} \int_0^{D/J} dx\, \left\langle \delta
\big( x - F(T_K) \big) \right\rangle \right|,
\end{equation}
where 
\begin{equation}
\label{eq:FTK}
F(T_K)= \frac{1}{2} \sum_{n=1}^N \frac{x_n}{s_n} \tanh \left(  
\frac{s_n}{2 \kappa} \right). \nonumber
\end{equation}
We note that a similar expression appears in the calculation of the
distribution of NMR-Knight shifts for metallic grains with a finite
level width $\Gamma$.\cite{pe93,b94,efetov} There, an exact solution
of the problem was derived within RMT in the limit $N \gg 1$. However,
two aspects make the problem defined by Eq. (\ref{eq:ptk}) much harder
to solve. First, in the NMR-Knight shift case, instead of
Eq. (\ref{eq:FTK}), the calculation involves a sum over terms which
behave as $1/(s_n^2 + \Gamma^2)$, decaying faster than $1/s_n$ at
large energies. Second and more importantly, since the Kondo
temperature enters in Eq. (\ref{eq:FTK}) essentially as the low-energy
cutoff of the sum, we cannot apply the techniques used in
Refs. \onlinecite{pe93,b94,efetov} to the calculation of the
distribution of Kondo temperatures. Thus, the derivation of an exact,
closed expression for the $P(T_K)$ is very nontrivial, even in RMT.

In RMT, for $\langle T_K \rangle \gg \Delta$, one would expect the
distribution of Kondo temperatures to be close to a Gaussian. The
reasoning behind that is the following. In this limit one can go back
to Eq. (\ref{eq:sctk}), expand the $\tanh$ factor and find that $T_K$
is given approximately by a sum of random variables (the wave function
amplitudes). The central-limit theorem then ensures that a Gaussian
distributed variable results from summing over independent random
variables. However, as we show below, even for large $\langle T_K
\rangle$ one finds a non-Gaussian behavior at the tails of the
distribution. This feature is enhanced when considering disorder
effects beyond RMT due to the appearance of local correlations between
wave functions at different energies. It is therefore clear that in
order to obtain the correct distribution of the Kondo temperature in
disordered metals, it is crucial to abandon the assumption of
independent random wave functions in Eq. (\ref{eq:sctk}). The main
effect of wave function correlations is to broaden the distribution of
$T_K$. This broadening survives the large-volume limit. In
Sec. \ref{sec:numerics} we show that correlations also induce a
bimodal structure in the distribution. This should be compared with
the rather featureless distribution obtained when one uses
uncorrelated RMT wave functions amplitudes to evaluate
$T_K$.\cite{jetp}

For reference, the behavior of the Kondo temperature in the clean
limit is discussed in Appendix \ref{sec:clean}.

\section{Fluctuations of $T_K$ in a disordered metal}
\label{sec:disorderedmetals}

The effect of nonmagnetic disorder on the statistical fluctuations of
the Kondo temperature is a difficult problem to study
analytically. However, in the weak-disorder limit it is possible to
derive expressions for the average and standard deviations of $T_K$
{\it beyond RMT} by taking explicitly into account wave function
correlations. To this end, we recall that the LDOS is defined as
\begin{equation} 
\label{eq:LDOS}
\rho ( {\bf r},E) = \sum_{n=1}^N |\psi_n ( {\bf r}) |^2 \delta \left(
E - E_n \right).
\end{equation} 
In our calculations, we will make use of the well-known correlation
function of LDOS, as defined by
\begin{eqnarray}
\label{eq:R2}
R_2({\bf r}, \omega) & = & \left\langle \rho({\bf r},E)\, \rho ({\bf
r}, E + \omega) \right\rangle \nonumber \\ & = & \frac{1}{\nu^2}
\sum_{n,m} \left\langle | \psi_n({\bf r})|^2\, |\psi_m ({\bf r})|^2\,
\delta (E-E_n) \right. \nonumber \\ & & \left. \times \delta (E
+\omega- E_m) \right\rangle.
\end{eqnarray}
It is convenient to separate diagonal from off-diagonal terms in
Eq. (\ref{eq:R2}). The correlation function of LDOS can then be
rewritten in terms of the spectral correlation function $R_2(\omega)$,
\begin{eqnarray}
\label{eq:R2aux}
R_2({\bf r}, \omega) & = & R_2 (\omega)\, \Omega^2 \left\langle
|\psi_n({\bf r})|^2\, |\psi_m ({\bf r})|^2 \right\rangle \mid_{\omega
  = E_n - E_m} \nonumber \\ & & +\, \delta (\omega/\Delta)\, \Omega^2
\left\langle |\psi_n({\bf r})|^4 \right\rangle.
\end{eqnarray}
Let us introduce the standard dimensionless parameter $g = E_F \tau$
to quantify the disorder strength. For $\omega<E_c$, the function $R_2
(\omega)$ is an oscillatory decaying function with corrections of
order $1/g^2$ to the leading RMT term. For the time-reversal symmetric
case ($\beta=1$), it reads
\begin{eqnarray}
R_2 \left( s=\frac{\omega}{\Delta} \right) &=& 1 - \frac{\sin^2 (\pi
s)}{\pi^2 s^2} \nonumber \\ &-& \left[ \frac{\pi}{2} \mbox{sgn} (s) -
\mbox{Si} (\pi s) \right] \left[ \frac{\cos(\pi s)}{\pi s} -
\frac{\sin(\pi s)}{(\pi s)^2} \right] \nonumber \\ &+& O(1/g^2),
\end{eqnarray}
where, $\mbox{Si}(z) = \int_0^z d y \sin y/y$.\cite{mirlin} For
frequencies exceeding the Thouless energy, $\omega>E_c$, the
oscillatory part of the spectral correlation function decays
exponentially, while there is a correction of order $1/g^2$ which
decays as $1/s$ without oscillations. While for pure RMT there are no
correlations between wave functions at different energies, in systems
with white-noise disorder these correlations are of order $1/g$,
namely,\cite{mirlin}
\begin{equation}
\label{psilk}
\Omega^2 \left\langle |\psi_{n}({\bf r})|^2 |\psi_{m} ({\bf r})|
\right\rangle_{\omega=E_n-E_m} = 1 + \frac{2}{\beta} {\rm Re}\,
\Pi(\omega)
\end{equation}
for $n \neq m$, while
\begin{equation}
\Omega^2 \left\langle |\psi_n({\bf r})|^4 \right\rangle = \left(
 1+\frac{2}{\beta} \right) \left[ 1 + \frac{2}{\beta} {\rm Re}\,
 \Pi(0) \right].
\end{equation}
These analytical results have recently been confirmed numerically in
Ref. \onlinecite{mub}. Note that the correlation is stronger when
time-reversal symmetry is present ($\beta =1$). The dependence on the
disorder enters through the summation over diffuson modes, which is
here represented by
\begin{equation} \label{dpropagator}
\Pi(\omega) = \frac{\Delta}{\pi} \sum_{{\bf q}}\frac{1}{D_e {\bf q}^2
- i \omega}.
\end{equation}
In two dimensions (2D) and for $L>l$, one obtains\cite{mirlin}
\begin{equation}
\label{eq:Pi2D}
\Pi^{(2D)}(\omega) = \frac{1}{2 \pi g} \ln \left( \frac{1/l^2 - i
\omega/D_e}{1/L^2 -i \omega/D_e} \right),
\end{equation}
with $l = v_F \tau$ and $D_e = v_Fl/2.$

Since the Kondo temperature is defined by a sum over all eigenstates
in the band, fluctuations of wave function amplitudes at the position
of the magnetic impurity add up quite effectively when the wave
functions are correlated over a large energy range. These correlations
are nonzero over an interval of the order of the elastic scattering
rate $1/\tau$. Thus, one can expect that fluctuations of the Kondo
temperature can exist even in the thermodynamic limit, when both
$\Delta$ and $E_c$ vanish.

In order to explicitly connect the Kondo temperature to the LDOS
fluctuations, we define $T_K^{(0)}$ as the Kondo temperature for a
non-fluctuating LDOS, $\nu$, and use Eq. (\ref{eq:LDOS}) into
(\ref{eq:tk}) to find\cite{zu96}
\begin{equation}
\label{eq:TK_transcd}
T_K = T_K^{(0)} \exp \left[ \int_{-E_F}^{D-E_F} dE\, \frac{\delta
\rho({\bf r}, E+E_F)}{2 \nu E} \tanh \left( \frac{E}{2 T_K} \right)
\right],
\end{equation}
where $\delta \rho \equiv \rho - \nu$. From the fact that, on average,
the deviations of the local density of states vanishes, $\langle
\delta \rho\rangle =0$, and by expanding the right-hand side of
Eq. (\ref{eq:TK_transcd}) to second order in $\delta \rho$, we find
\begin{eqnarray} 
\label{lntk}
\left\langle \ln^2 \left( \frac{T_K}{T_K^{(0)}} \right) \right\rangle
& = & \int_{-E_F}^{D-E_F} dE\, \int_{-E_F}^{D-E_F} dE' \nonumber \\ &
& \times \left\langle \tanh \left( \frac{E}{2 T_K^{(0)}} \right) \tanh
\left( \frac{E'}{2 T_K^{(0)}} \right) \right.\nonumber \\ & & \times
\left. \frac{\delta \rho({\bf r}, E_F + E)}{2 \nu E} \frac{\delta
\rho({\bf r}, E_F + E')}{2 \nu E'} \right\rangle. \nonumber \\
\end{eqnarray}
Note that Eq. (\ref{lntk}) provides an approximate expression for
standard deviation of the Kondo temperature, $\delta T_K$: When
$\delta T_K \ll \langle T_K \rangle$, we have $\delta T_K \approx
T_K^{(0)} \sqrt{\left\langle \ln^2 \left( T_K/T_K^{(0)} \right)
\right\rangle}$.

Using Eq. (\ref{eq:R2aux}), the right-hand side of Eq. (\ref{lntk})
can be broken down into three contributions, namely,
\begin{eqnarray}
\label{eq:lnTK_g}
\left\langle \ln^2 \left( \frac{T_K}{T_K^{(0)}} \right) \right\rangle
& = & 2 \left[ S_{\beta} \left( \tau, \Delta, T_K^{(0)} \right) + V_{\beta}
\left( \tau, E_c, T_K^{(0)} \right) \right. \nonumber \\ & & \left. +\
Q_{\beta} \left( \tau, E_c, T_K^{(0)} \right) \right].
\end{eqnarray}
The function $S_{\beta}$ arises from the spectral self-correlation
term $\delta(\omega/\Delta)$ and is given by
\begin{eqnarray}
S_{\beta} (\tau,\Delta,T_K) & = & \frac{\Delta}{8} \left(1 +
\frac{2}{\beta} \right) \left[1 + \frac{2}{\beta} {\rm Re}\, \Pi(0)
\right] \nonumber \\ && \times \int_{-E_F}^{D-E_F} \frac{dE}{E^2}
\tanh^2 \left( \frac{E}{2 T_K} \right).
\end{eqnarray}
This term is of order $\Delta/T_K$ and therefore vanishes in the
thermodynamic limit as $\Delta \rightarrow 0$. The decaying part of
the spectral correlation function yields the second term on the
right-hand side of Eq. (\ref{eq:lnTK_g}),
\begin{eqnarray}
V_{\beta} (\tau, E_c, T_K) & = & \frac{1}{8} \int \int_{-E_F}^{D-E_F}
\frac{dE\, dE^\prime}{E\, E^\prime}\, \tanh \left( \frac{E}{2 T_K}
\right) \nonumber \\ && \times \tanh \left( \frac{E^\prime}{2
T_K}\right)\, \left[ R_2 (E-E^\prime) -1 \right] \nonumber \\ &&
\times \left[ 1+ \frac{2}{\beta} {\rm Re}\, \Pi (E -E^\prime)\right].
\end{eqnarray}
In 2D, $R_2 (\omega)-1$ decays as $1/\omega$ for frequencies exceeding
the Thouless energy. This causes $V_{\beta}$ to be nonzero as $E_c
\rightarrow 0$, but this term turns out to be only of order
$1/g^2$. If correlations of wave functions are taken into account only
up to first order in $1/g$, $V_\beta$ can be discarded. At finite
$E_c$, however, it needs to be taken into account since it yields
terms of order $E_c/T_K$ due to spectral correlations at small
frequencies, $\omega<E_c$.

The term arising purely from the correlations of wave functions is
given by
\begin{eqnarray}
\label{eq:Qfinal}
Q_{\beta} (\tau, E_c, T_K) & = & \frac{1}{4 \beta } \int
\int_{-E_F}^{D-E_F} \frac{ dE\, dE^\prime}{E\, E^\prime}\, \tanh
\left( \frac{E}{2 T_K} \right) \nonumber \\ && \times \tanh \left(
\frac{E^\prime}{2 T_K}\right)\, {\rm Re}\, \Pi (E -E^\prime).
\end{eqnarray}
This term survives the thermodynamic limit. For instance, let us
consider the 2D case. From Eq. (\ref{eq:Pi2D}), we find
\begin{equation}
\label{eq:Pol2d}
{\rm Re}\, \Pi^{(2D)}(\omega) = \frac{1}{4 \pi g} \ln \left( \frac{
1/4\tau^2 + \omega^2} {E_c^2 + \omega^2}\right).
\end{equation}
For $1/\tau \gg T_K$, one can use Eq. (\ref{eq:Pol2d}) with $E_c=0$ to
write the double integral in Eq. (\ref{eq:Qfinal}) in terms of
polylogarithm functions. To leading order and after setting $E_F =
D/2$, we obtain
\begin{equation}
\label{eq:rbeta}
Q_{\beta}^{(2D)} (\tau, 0,T_K) \approx \frac{1}{ 6\pi\, g\, \beta}
\left[ \ln \left( \frac{E_F}{g\, T_K} \right) \right]^3,
\end{equation}
which is valid when $E_F/T_K \gg g \gg 1$. Combining
Eqs. (\ref{eq:lnTK_g}) and (\ref{eq:rbeta}), we arrive at
\begin{equation}
\label{eq:lnTK}
\left\langle \ln^2 \left( \frac{T_K}{T_K^{(0)}} \right)
\right\rangle_{2D} \approx \frac{1}{3\pi\, g\, \beta }\, \left[ \ln
\left( \frac{1}{\tau\, T_K^{(0)}} \right) \right]^{3},
\end{equation}
which in entirely {\it volume independent}.

Expressions for the $\langle T_K \rangle$ and $\delta T_K$ can also be
derived from Eq. (\ref{lntk}). However, in order to keep results
accurate up to second order in $\delta \rho$, one has to take into
account the correlation between $T_K$ and $\delta \rho$ when expanding
the right-hand side of Eq. (\ref{eq:TK_transcd}). Doing so, we obtain for the
average
\begin{eqnarray}
\label{eq:TK_g}
\langle T_K \rangle & = & T_K^{(0)} \left\{ 1+ \left[ 1+ T_K^{(0)}
\partial_{T_K^{(0)}} \right] \left[ S_{\beta} \left( \tau, \Delta,
T_K^{(0)} \right) \right. \right. \nonumber \\ & & \left. \left. +
V_{\beta} \left( \tau, E_c, T_K^{(0)} \right) + Q_{\beta} \left( \tau,
E_c, T_K^{(0)} \right) \right] \right\}.
\end{eqnarray}
Similarly, for the standard deviation, we arrive at
\begin{eqnarray}
\label{eq:dTK_g}
\left( \delta T_K \right)^2 & = & 2\, \left( T_K^{(0)} \right)^2
\left[ S_{\beta} \left( \tau, \Delta, T_K^{(0)} \right) + V_{\beta}
\left( \tau, E_c, T_K^{(0)} \right) \right. \nonumber \\ & & \left.  +
Q_{\beta} \left( \tau, E_c, T_K^{(0)} \right) \right].
\end{eqnarray}
Combining Eq. (\ref{eq:lnTK}) with (\ref{eq:dTK_g}) we finally get
\begin{equation}
\label{2dwidth}
\left. \delta T_K \right|_{2D} \approx T_K^{(0)} \frac{1}{\sqrt{3\pi g
\beta}}\, \left[ \ln \left( \frac{1}{\tau\, T_K^{(0)}} \right)
\right]^{3/2}.
\end{equation}
This expression shows that even in the weak-disorder metallic regime
($g \gg 1 $) the width of the distribution of Kondo temperatures
remains finite as $\Delta$ and $E_c$ go to zero. It is interesting to
note that this width is reduced by a factor $1/\sqrt{2}$ when
time-reversal symmetry is broken ($\beta=2$). This effect can be
understood if we recall that the weak-localization corrections, which
account for exactly half of the wave function correlation correction,
become suppressed when time-reversal symmetry is fully broken. Then,
only wave function correlations due to the classical diffusion
propagator remain.

It is also important to remark that Eq. (\ref{2dwidth}) is valid
despite the fact that all electron states have a finite localization
length $\xi$ in dimensions not exceeding two. In particular, we note
that the width Eq. (\ref{2dwidth}) exceeds previous estimates, such as
$\delta T_K \sim \sqrt{\Delta_{\xi} T_K^{(0)}}$, which were based on
the application of RMT on the scale of the local level spacing
$\Delta_{\xi}$ and disregarded the dominant effect of wave function
correlations that we consider here.\cite{moriond,grempel}

From Eqs. (\ref{eq:lnTK}) and (\ref{eq:TK_g}), in the limit $1/\tau
\gg T_K$, the average Kondo temperature is found to increase with the
disorder strength as
\begin{equation}
\label{eq:TK2D}
\langle T_K \rangle_{2D} \approx T_K^{(0)}\left\{ 1+ \frac{1}{ 6\pi g
\beta }\, \left[ \ln \left( \frac{1}{\tau\, T_K^{(0)}} \right)
\right]^{3} \right\}.
\end{equation}
Although the disorder-induced term in Eq. (\ref{eq:TK2D}) is only of
order $1/g$, it is enhanced by the large factor coming from the third
power of the logarithm of $T_K^{(0)} \tau$.

These calculations can be repeated using the diffusion propagator for
quasi-one-dimensional (Q1D) wires. For open boundary conditions, the
diffusion propagator in Eq. (\ref{dpropagator}) becomes
\begin{equation}
{\rm Re}\, \Pi^{(Q1D)} (\omega ) = \left\{ \begin{array}{lr} L/(k_F^2
l\, A), & \omega \ll E_c \\ \sqrt{6}/(k_F^2 A\, \sqrt{\tau \omega}), &
E_c < \omega < 1/\tau, \\ 0, & \omega \gg 1/\tau,
\end{array} \right.
\end{equation}
where $L$ is the length and $A$ is the cross sectional area of the
quasi-1D wire. Therefore, in the long-wire limit, with $E_c \ll T_K
\ll 1/\tau$, we find
\begin{equation}
\label{eq:lnTKQ1D}
\left\langle \ln^2 \left( \frac{T_K}{T_K^{(0)}} \right)
\right\rangle_{Q1D} = \frac{4 \pi \sqrt{3}}{\beta} \frac{1}{k_F^2 A}
\frac{1}{\sqrt{\tau T_K^{(0)}}},
\end{equation}
independently of the position of the Fermi energy. Hence, the average
Kondo temperature increases with disorder by an inverse square root
term,
\begin{equation}
\label{eq:TK_gQ1D}
\langle T_K \rangle_{Q1D} \approx T_K^{(0)}\, \left( 1 + \frac{ \pi
\sqrt{3}}{\beta} \frac{1}{k_F^2 A} \frac{1}{\sqrt{\tau T_K^{(0)}}}
\right),
\end{equation}
while the variance is nonvanishing by the square root of the same
factor,
\begin{equation}
\label{eq:dTK_gQ1D}
\left. \delta T_K \right|_{Q1D} \approx T_K^{(0)} \, \sqrt{ \frac{4
\pi \sqrt{3}}{\beta} \frac{1}{k_F^2 A} \frac{1}{\sqrt{\tau T_K^{(0)}
}}}.
\end{equation}
The latter result is in agreement with the $1/g^{1/4}$ dependence
reported in Ref. \onlinecite{micklitz05}, where the asymptotic
behavior for the 2D case, $(J^3/g)^{1/2}$, Eq. (\ref{2dwidth}) was
also found.

\section{Numerical studies of a disordered model}
\label{sec:numerics}

The analytical results presented in Sec. \ref{sec:disorderedmetals}
 are applicable only in the weak-disorder regime and only
predict the behavior of the two lowest moments of the distribution
of Kondo temperatures. In order to study more thoroughly the
statistical fluctuations of the Kondo temperature, as well as to check
the accuracy of Eqs. (\ref{eq:TK_g}), (\ref{eq:dTK_g}),
(\ref{eq:TK_gQ1D}), and (\ref{eq:dTK_gQ1D}), we have carried out
numerical studies of the fluctuations of $T_K$ using a tight-binding
model with nearest-neighbor hopping and random site potential to
describe the conduction electrons. The model Hamiltonian reads
\begin{equation}
\label{eq:Htb}
H = -t \sum_{\langle ij \rangle} \left( c_{i}^\dagger c_{j} + {\rm
h.c.} \right) + \sum_{i=1}^N V_i\, c_{i}^\dagger c_{i},
\end{equation}
where each site potential $V_i$ is drawn from a flat box distribution
of width $W$ centered at zero. We assume each eigenstate of $H$ to be
spin degenerate, therefore sums over spins are implicit in
Eq. (\ref{eq:Htb}). We consider only square lattices. Aspect ratios
and boundary conditions are as follows: To simulate two-dimensional
systems, we use a square geometry with periodic boundary conditions
along both directions; For quasi-one-dimensional systems, we adopt a
rectangular geometry with hard-wall (periodic) boundary conditions
along the shortest (longest) length.

Employing standard numerical techniques, we have diagonalized the
Hamiltonian $H$ for a large set of realizations of the disordered
potential and different lattice geometries. The resulting
eigenenergies $E_n$ and eigenvectors $\psi_n(i)$, $n=1,\ldots,N$, were
used in conjunction with Eq. (\ref{eq:tk}) to determine $T_K$. In all
simulations the Fermi level was placed at the lower quarter of the
band in order to avoid the large peak in the density of states at
$E=0$, reminiscent of the van Hove singularity found in the clean
limit. No unfolding of energy levels was used. The simulations were
carried out for $20\times 20$, $10\times 100$, and $8\times 200$
lattice sizes, with 1,000 realizations for each value of $W$. In order
to increase statistics, a total of 36 ($20 \times 20$), 196 ($10
\times 100$), and 297 ($8 \times 200$) different magnetic impurity
sites were used. No magnetic flux was included, thus all results refer
to the time-reversal symmetric class. Partial results for the
$20\times 20$ case were previously reported in Ref. \onlinecite{jetp}.

\begin{figure}[t]
\includegraphics[width=7.8cm]{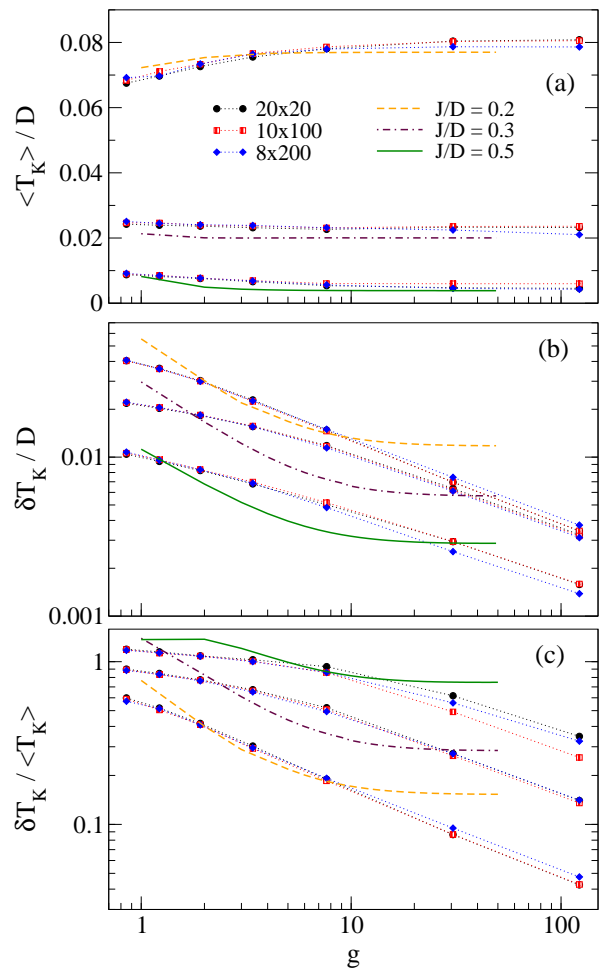}
\caption{(Color on-line) Average (a), standard deviation (b), and
their ratio (c) as a function of disorder strength for three values of
$J/D$ and three lattice geometries: two-dimensional ($20 \times 20$)
and quasi-one-dimensional wires ($10 \times 100$ and $8 \times
200$). Dotted lines are only guides to the eyes. The dashed,
dotted-dashed, and solid curves are the analytical prediction for
$J/D=0.20$, 0.30, and 0.50, respectively, based on
Eqs. (\ref{eq:TK_g}) and (\ref{eq:dTK_g}) and taking into account the
finite values of $E_c$ and $\Delta$. The analytical curves were
plotted only within their region of validity. The same horizontal
scale is used in (a), (b), and (c).}
\label{fig:stat_g}
\end{figure}

A connection between the tight-binding disorder strength $W$ and the
dimensionless parameter $g$ (see Sec. \ref{sec:disorderedmetals}) can
be made by recalling the expression for the nonmagnetic scattering
rate in Born approximation,
\begin{equation}
\label{eq:tau}
\frac{1}{\tau} = 2 \pi\, \nu\, \left\langle V^2 \right\rangle,
\end{equation}
where $\nu = 1/\Delta\, L^2$ (we set the lattice constant $a=1$,
therefore $N=L^2$). Noting that $\left\langle V^2 \right\rangle =
W^2/12$ and $\Delta = D/N$, we obtain
\begin{equation}
\label{eq:tau2}
\frac{1}{\tau} = \frac{\pi}{6} \frac{W^2}{D}.
\end{equation}
Recalling that $g = E_F\,\tau$ and setting $E_F = D/4$, we arrive at
\begin{equation}
\label{eq:g}
g = \frac{3}{2\pi} \left( \frac{D}{W} \right)^2.
\end{equation}
Equation (\ref{eq:tau}) is valid only in the perturbative sense, when
disorder is not too strong, and should break down near
$g=1$. Furthermore, when deriving Eq. (\ref{eq:tau2}) we assumed a
flat density of states (i.e., a parabolic dispersion relation), which
is a rough approximation. Thus, taking into account such limitations,
$g$ has to be considered merely as a convenient way to parametrize the
transition from a metallic ($g \gg 1$) to a strongly disordered,
localized regime ($g <1$). It is important to remark that $g$ is not
equal to the dimensionless conductance parameter ${\cal G} \equiv
E_c/\Delta$.

\subsection{Average and standard deviation of $T_K$}
\label{sec:momentsTK}

The dependence on $g$ of the average and standard deviation of the
Kondo temperature, as well as their ratio, is shown in
Fig. \ref{fig:stat_g} for several values of the exchange coupling
$J$. The solid, dashed, and dotted-dashed curves are the analytical
predictions for a $20 \times 20$ (two-dimensional) lattice based on
Eqs. (\ref{eq:TK_g}) and (\ref{eq:dTK_g}) and taking into account the
finite values of $E_c$ and $\Delta$. The clean-limit value of the
Kondo temperature, $T_K^{(0)}$, was obtained from
Eq. (\ref{eq:tkusual}) with the proper adjustment for the Fermi level
position.

Figure \ref{fig:stat_g} contains several important features. First,
the plots show that the statistics of $T_K$ is quite independent of
lattice size and geometry. Even wires with an aspect ratio of 1:25
still present statistical fluctuations of $T_K$ consistent with a
two-dimensional geometry after the appropriate rescaling. Deviations
are only apparent for the largest value of $g$ considered. This
behavior can be understood if we recall that for diffusive dynamics
the crossover from 2D to the Q1D regime occurs when the elastic mean
free path becomes larger than the wire width but remains smaller than
the wire length ($L_y < l < L_x$). For a square lattice with weak site
disorder ($g \gg 1)$, assuming a flat density of states, it is
straightforward to show that
$$ 
\frac{l}{a} \approx g\, \sqrt{ \frac{D}{2E_F}}.
$$ 
Applying this expression to the $10 \times 100$ and $8 \times 200$
lattices we find that the 2D-Q1D crossover happens around $g = g_c
\approx 7$ and 6, respectively. Thus, at $g < g_c$, the electronic
eigenstates should follow the statistics of a two-dimensional
disordered lattice, irrespective of the wire aspect ratio, and this is
indeed what we observe. At $g < 1$, all states are localized and
therefore the statistics of $T_K$ becomes independent of lattice size,
geometry, or boundary conditions.

In order to reach a regime where the analytical expressions for Q1D
[Eqs. (\ref{eq:TK_gQ1D}) and (\ref{eq:dTK_gQ1D})] are applicable, we
would need narrower wires or much weaker disorder. At the same time,
we would need to maintain a large number of transversal modes and
longitudinal diffusion. In practice, these constraints can only be
satisfied for much larger lattices than those we have investigated.

It is also clear from Fig. \ref{fig:stat_g} that the average Kondo
temperature is only substantially affected by disorder when
localization sets in ($g\rightarrow 1$). This behavior is reminiscent
of the independence of the critical temperature on nonmagnetic
disorder in weakly disordered conventional
superconductors.\cite{rickayzen} In the weak-disorder regime, $\langle
T_K \rangle$ follows closely the bulk clean case value of the Kondo
temperature [see Eq. (\ref{eq:tkusual})], which is consistent with the
prediction of Eq. (\ref{eq:TK_g}). Deviations only become large in the
strong disordered regime, where the analytical expressions are not
expected to be valid. The situation is less satisfactory for $\delta
T_K$, where the deviations from the analytical curves are seen for all
disorder strengths. Note that the agreement with Eq. (\ref{eq:dTK_g})
is better for large values of $J/D$.

Overall, the analytical expressions do capture the qualitative aspects
correctly. There is an enhancement of the Kondo temperature and an
increase of its variance as the disorder strength increases and $g$
decreases. According to the analytical calculations, this is mainly
due to the appearance of weak correlations of wave functions at
different energies.


\begin{figure*}[t]
\includegraphics[width=15cm]{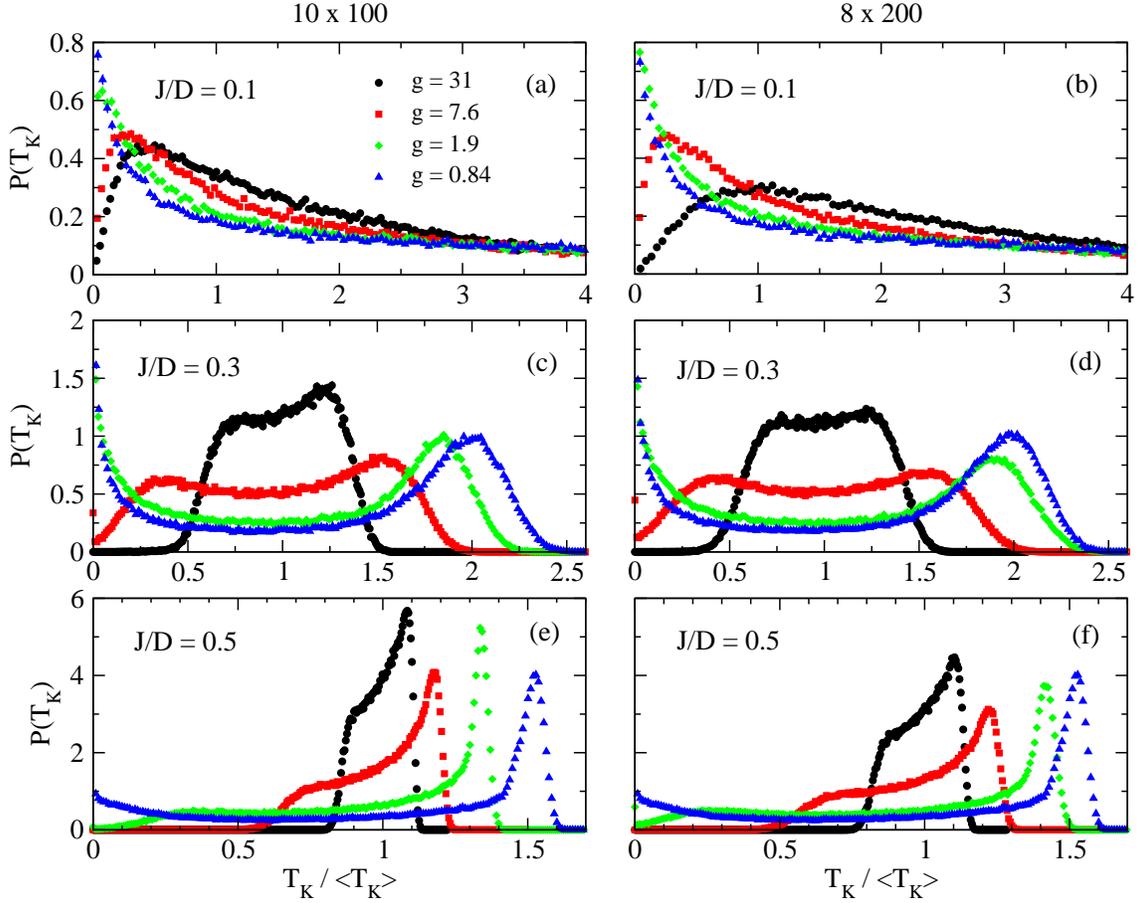}
\caption{(Color on-line) The distribution of the Kondo temperatures
for $10\times 100$ and $8 \times 200$ lattices for a range of disorder
strengths and three different values of $J/D$. Events corresponding to
$T_K=0$ fall outside the vertical scale in (a) and (b).}
\label{fig:ptk_comp}
\end{figure*}


\subsection{The distribution of $T_K$}
\label{sec:distribution}

The effect of disorder is much more drastic for the distribution of
Kondo temperatures as a whole. In Ref. \onlinecite{jetp}, it was found
that a bimodal structure appears even for weak disorder and large
exchange coupling. To further investigate this effect and its
dependence on sample geometry, we present in Fig. \ref{fig:ptk_comp}
the distribution of $T_K$ obtained from numerical calculations
involving $10 \times 100$ and $8 \times 200$ lattices. These should be
compared with Fig. 2 in Ref. \onlinecite{jetp}, where the $20 \times
20$ case was considered.

Two distinct peaks are clearly visible when $J/D$ is not too small. As
$J/D$ increases, weight is transfered from the low-$T_K$ to the
high-$T_K$ peak. The distributions never come as close to a Gaussian
as they do in the RMT case (see Ref. \onlinecite{jetp} for the RMT
distributions). The distinct features of the distributions are a wide
plateau in $P(T_K)$ for weak disorder and a marked second peak at
small $T_K$ for strong disorder. Since Eq. (\ref{eq:rbeta}) indicates
that the enhancement of fluctuations in the Kondo temperature survive
the thermodynamic limit, we believe that the overall form observed in
Fig. \ref{fig:ptk_comp} should persist as the linear size of the
lattice is increased. However, to verify this conjecture, one would need
to perform a finite-size scaling analysis over a much wider range of
lattice sizes. For an analytical proof, at least a few moments higher
than the variance would have to be calculated. Both are beyond the
scope of this paper. The finite-size rescaling will be investigated in
a future work using a more refined numerical technique.\cite{future}

It is important to remark that a finite-size scaling analysis was
performed in Ref. \onlinecite{grempel} in the {\it strong disorder}
limit only. It was found that a critical feature seen in
Figs. \ref{fig:ptk_comp}(c)-(f), namely, the divergence of the
distribution as $T_K \rightarrow 0$ at strong disorder, does survive
finite-size scaling. In Ref. \onlinecite{grempel}, this divergence was
found numerically to be a power law and it was argued that the
exponent depended solely on the lattice space dimension. However, no
analytical derivation was given to justify such a behavior. This issue
is addressed in Sec. \ref{sec:strongdisorder}.

\section{Distribution of $T_K$ in the localized regime}
\label{sec:strongdisorder}

For the purpose of investigating the power-law behavior reported in
Ref. \onlinecite{grempel}, let us consider the localized regime in
more detail. When the localization lengths $\xi_n$ of all states
within the energy band are smaller than the system size, one can
calculate the distribution function analytically by noting that only
states which are located within a distance smaller than $\xi_n$ to the
magnetic impurity state contribute significantly to the Kondo
screening. The probability density of the $n$th eigenstate at the
position of the magnetic impurity, ${\bf r}_0 =0$, is then given by
$(L/\xi_n)^d \exp( -2|{\bf r}|/\xi_n)$. We can then average over the
center-of-mass positions ${\bf r}_n$ of the states closest in energy
to the Fermi level. Note that localized states can overlap. When the
localization length is sufficiently large, several states can be found
within a localization volume $\xi^d$ and the corresponding energy
levels will obey Wigner-Dyson-statistics. In particular, the two
energy levels closest to the Fermi energy repel each other on the
scale $\Delta_{\xi} =1/(\nu \xi^d)$. Under these assumptions, we can
use Eq. (\ref{eq:ptk}) to derive a distribution of Kondo temperature
in the localized regime, namely,
\begin{equation}
P(T_K) = \frac{\Omega}{T_K}\, \left\langle F(\epsilon/T_K)
\right\rangle_{\epsilon},
\end{equation}
where $\Omega$ is the sample volume,
\begin{equation}
F(x) = \frac{x}{\tanh (x/2) \cosh^2 (x/2)} \ln^{d-1} \left[ \frac{x
T_K }{\Delta_{\xi} \tanh(x/2)} \frac{D}{J} \right],
\end{equation}
and $\langle \ldots \rangle_{\epsilon}$ denotes the average over the
level spacing $\epsilon$ between the states within the localization
volume $\xi^d$ which are close to the Fermi energy $E_F$. Let us now
look at both ``weak'' and ``strong'' localization regimes.

For large localization lengths, $\Delta_{\xi}$ is much smaller than
the bandwidth and we can assume that energy levels in the vicinity of
the magnetic impurity repel within the energy window
$\Delta_{\xi}$. As a result, for Kondo temperatures exceeding the
local level spacing, $T_K > \Delta_{\xi}$, the distribution becomes
\begin{equation}
P(T_K)\mid_{T_K > \Delta_{\xi}} = \frac{\Omega}{T_K} \ln^{d-1} \left(
\frac{T_K}{\Delta_{\xi}}\right),
\end{equation}
which indeed contains a power law divergence at small $T_K$ but with
an exponent independent of the space dimension. As $T_K$ becomes
smaller than $\Delta_{\xi}$, this divergence is cutoff due to level
repulsion. For a fixed level spacing $\epsilon = \Delta_{\xi}$, one
gets an exponential cutoff $P(T_K)\mid_{T_K < \Delta_{\xi}} \sim
(\Delta_{\xi}/T_K^2) \exp (-\Delta_{\xi}/T_K)$. However, the spacing
$\epsilon$ is known to follow the Wigner surmise, $P(s=
\epsilon/\Delta_{\xi}) = a_{\beta}\, s^{\beta}\, \exp (- c_{\beta}\,
s^2)$, where, $a_{\beta} = \pi/2$ ($32/\pi$) and $c_{\beta} = \pi/4$
($4/\pi$), for $\beta =1$ (2). Performing the average over $\epsilon$,
we arrive at
\begin{equation}
P(T_K)\mid_{T_K < \Delta_{\xi}} = \frac{\Omega\, a_{\beta}}{\beta+2}
\frac{T_K^{\beta}}{\Delta_{\xi}^{\beta+1}}.
\end{equation}
This distribution decays (rather than diverge) for $T_K <
\Delta_{\xi}$.

In the strongly localized regime, $g \ll 1$, $\xi$ becomes of the
order of the Fermi wave length $\lambda_F$. Then, only a single energy
level has appreciable overlap with the magnetic impurity. The spacing
between the energy level and the Fermi energy obeys a Poissonian
distribution, $ P(\epsilon) =
\exp(-|\epsilon|/\Delta_{\xi})$. Averaging over $\epsilon$ yields
\begin{equation}
P(T_K)\mid_{T_K > \Delta_{\xi}} \sim \frac{1}{T_K} \ln^{d-1} \left(
\frac{T_K}{\Delta_{\xi}}\right)
\end{equation}
and
\begin{equation}
P(T_K)\mid_{T_K < \Delta_{\xi}} \sim \frac{\Delta_{\xi}/T_K^2}{(2 +
\Delta_{\xi}/T_K)^2}\, \ln^{d-1} \left( \frac{D}{J} \right).
\end{equation}
Thus, we also find a power law increase of $P(T_K)$ with decreasing
$T_K$ in the strong-disorder limit. But this divergence is also cutoff
at $T_K < \Delta_{\xi}$ and $P(T_K)$ converges to a finite value:
$P(T_K) \mid_{T_K \ll \Delta_{\xi}} \rightarrow 1/\Delta_{\xi}$.
  
In summary, in contrast with the interpretation of the numerical
scaling results in Ref. (\onlinecite{grempel}), we find that the power
of the divergence of $P(T_K)$ does not depend on space dimension, but
is rather equal to 1. However, in both cases, the divergence is cut
off at $T_K < \Delta_{\xi}$: For $\xi \gg \lambda_F$, there is a
power-law decay, $P(T_K) \sim T_K^{\beta}$, while for $\xi \rightarrow
\lambda_F$ it approaches a constant value, $P(T_K) \mid_{T_K \ll
\Delta_{\xi}} \rightarrow 1/\Delta_{\xi}$. These analytical results
are found to be in qualitative agreement with our numerical results
[see Fig. \ref{fig:ptk_comp}].
  
It should be noted that while the events corresponding to $T_K=0$ are
not always shown in Fig. \ref{fig:ptk_comp}, they were taken into
account in the evaluation of $\langle T_K \rangle$ and $\delta T_K$ in
Figs. \ref{fig:stat_g}. The long tail toward the small-$T_K$ region
indicates that disorder enhances the probability of having unscreened,
free magnetic moments at zero temperature. This is in agreement with
the data present in Fig. 4 of Ref. \onlinecite{jetp}, where the
probability of finding no solution to Eq. (\ref{eq:tk}) was plotted as
a function of $J/D$.

\section{Disordered grains: Random Matrix Theory}
\label{sec:rmt}

In a small, weakly disordered, phase-coherent metallic sample, such as
a metal grain or nanoparticle,\cite{mucciolo06} the single-particle
energy levels repel each other and the wave function intensities are
distributed randomly. This behavior becomes very relevant to several
physical properties of the system at low temperatures, namely, when
$T<E_c$ (see Fig. \ref{fig:kondoboxmap}). This in turn will affect the
Kondo temperature associate to magnetic impurities located inside the
sample.

For weak disorder, the dimensionless conductance parameter is very
large: ${\cal G} = E_c/\Delta \rightarrow \infty$. In this regime, the
spacing between consecutive levels within a window of energy $\omega <
E_c$ obeys the Wigner-Dyson distribution of RMT.\cite{mehta} Moreover,
within a given symmetry class, the wave function intensities $x_n$ and
the rescaled eigenenergies $s_n$ fluctuate independently. The
quantities $x_n$, $n=1,\ldots,N$, are themselves uncorrelated and obey
the so-called Porter-Thomas distributions.\cite{efetov,mirlin} For the
Gaussian unitary ensemble (GUE - broken time-reversal symmetry class)
and for the Gaussian orthogonal ensemble (GOE - time-reversal
symmetric class) these distributions are given by
\begin{equation}
\label{eq:PT}
P(x_n) = \left\{ \begin{array}{lr} e^{-x_n}, & \mbox{GOE}, \\ 
e^{-x_n/2} / \sqrt{2 \pi x_n}, & \mbox{GUE}. \end{array} \right.
\end{equation}

In order to gain insight about the statistical properties of $T_K$
over a wide range of values for $J$, we have solved
Eq. (\ref{eq:sctk}) numerically using eigenenergies $s_n$ obtained by
diagonalizing GOE and GUE random matrices of size $N=500$. The
resulting semi-circular spectrum was unfolded into a flat band
following standard procedures and the Fermi energy was set to the
middle of the band ($E_F=0$). Instead of using the random matrix
eigenfunctions to get the local wave function intensities $x_n$, we
generated these quantities directly from the Porter-Thomas
distributions as given in Eq. (\ref{eq:PT}).\cite{obs2} A total of 500
random matrix realizations were used for each ensemble type. For each
realization (energy spectrum), we simulated different impurity
locations by drawing 250 values of $x_n$, thereby increasing
substantially the statistics of $T_K$ without having to perform a
large number of matrix diagonalizations. In Ref. \onlinecite{jetp},
the distribution of Kondo temperatures obtained in this way were shown
for several values of $J/D$. Here we provide a detailed analysis of
these and other numerical and analytical results.

\begin{figure}[t]
\includegraphics[width=8.5cm]{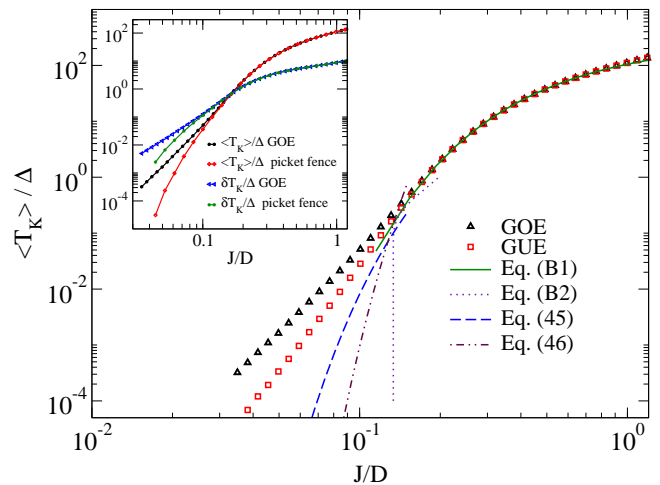}
\caption{(Color on-line) The dependence of average Kondo temperature
on the exchange coupling for the GOE and GUE compared with the
analytical solution in the clean case. Nonzero values of $T_K$ are
found below the clean-case threshold given by Eq. (\ref{jminus}) ($J_-
\approx 0.134$ here). In this region, $T_K$ is controlled by a few
levels close to the Fermi energy, as a comparison with the results of
the 2LS model, Eqs. (\ref{tlsO}) and (\ref{tlsU}), indicates. Inset:
The dependence of the Kondo temperature average $\langle T_K \rangle$
and standard deviation $\delta T_K$ on the exchange coupling $J$ for
unfolded GOE spectra as compared with those obtained for a spectrum of
equally spaced energy levels (picket fence). In both cases wave
function intensities were drawn from the GOE Porter-Thomas
distribution. It is only for $\langle T_K\rangle < \Delta$ that the
fluctuations of the energy levels result in an appreciable enhancement
of $\langle T_K \rangle$ and $\delta T_K$.}
\label{comparisonanalytic}
\end{figure}

In Fig. \ref{comparisonanalytic}, the GOE and GUE values of $\langle
T_K \rangle$ are plotted versus $J$ together with the curves obtained
in the clean, thermodynamic limit, namely, Eqs. (\ref{eq:tkusual}) and
(\ref{tksmall}). The average Kondo temperature is found to coincide
with the clean case regardless of the ensemble symmetry when $\langle
T_K \rangle > \Delta$. It is only for $\langle T_K \rangle < \Delta$
that it becomes dependent on the symmetry class. Note that spectral
fluctuations allows for solutions of Eq. (\ref{eq:sctk}) below the
clean-limit threshold value $J_-$ (see Appendix \ref{sec:clean}). We
will return to this point in Sec. \ref{sec:TLS}.

\subsection{Fluctuations in the large-$T_K$ limit ($T_K > \Delta$)}
\label{sec:largeTK}
   
The average and the standard deviation of the Kondo temperatures for
GOE are shown in the inset of Fig. \ref{comparisonanalytic}. They are
compared with results obtained for an equally spaced spectrum (picket
fence) where the eigenfunction intensities were also drawn from the
GOE Porter-Thomas distribution. It is clear that the average and
standard deviation of the Kondo temperature are insensitive to level
fluctuations when $T_K$ is larger than $\Delta$. We can use this fact
to obtain an approximate expression for the distribution of $T_K$ in
the RMT: Using Eq. (\ref{eq:PT}) and substituting the delta function
by its Fourier integral representation, we can perform the average
over the wave function amplitudes in Eq. (\ref{eq:ptk}) exactly. For
the orthogonal class, assuming an equally spaced spectrum, we obtain
\begin{equation}
\label{pfs}
P_{\rm GOE} (T_K) = \left| \frac{d}{dT_K} \sum_{r=1}^{N/2} \exp \left(
- \frac{D G_r}{J} \right) \prod_{n=1, n \neq r}^{N/2} \frac{G_n}{G_r
-G_n} \right|,
\end{equation}
where, 
\begin{equation}
G_n = (n-1/2) \coth \left( \frac{n-1/2}{2
\kappa}\right).
\end{equation}
Equation (\ref{pfs}) can only be further simplified in some limiting
cases, as outline below.

For $T_K \gg \Delta$, one can set $G_r = r-1/2$ for $r> \kappa$ and
$G_r = 0$ for $r < \kappa$. The Kondo distribution then gains a
stretched exponential form,\cite{jetp}
\begin{equation}
\label{eq:PTKlarge}
P_\beta \left( T_K \right) \sim \frac{1}{\Delta}\exp \left[
\beta\kappa \ln \left( \frac{e \kappa_0}{\kappa} \right)
-\beta\kappa_0 \right],
\end{equation}
where $\kappa_0 = T_K^{(0)}/\Delta$. Note that in the vicinity of
$T_K^{(0)}$ the distribution is close to a Gaussian: Near $T_K^{(0)}$
and for $\kappa_0 \gg 1$ it can be approximated as
\begin{equation}
\label{pgauss}
P_\beta\left( T_K \right) \sim \frac{1}{\Delta}\exp \left[ -\beta
(\kappa-\kappa_0)^2/2\kappa_0 \right].
\end{equation}
The departure from the Gaussian behavior occurs at the tails of the
distribution: For $\kappa \ll \kappa_0$ ($\kappa \gg \kappa_0$) the
curve defined by Eq. (\ref{eq:PTKlarge}) runs below (above) a
Gaussian. In Ref. \onlinecite{jetp}, it was noted that $P(T_K)$
becomes wider than Gaussian for small exchange couplings. Even for
larger $J$, when the average $\langle T_K \rangle$ exceeds $\Delta$,
clear deviations from the Gaussian behavior were found, with the
distribution showing asymmetric non-Gaussian tails. There was good
quantitative agreement between Eq. (\ref{eq:PTKlarge}) and the
numerical data where $\kappa_0>1$ and $J<D$.

\begin{figure}[t]
\includegraphics[width=7.5cm]{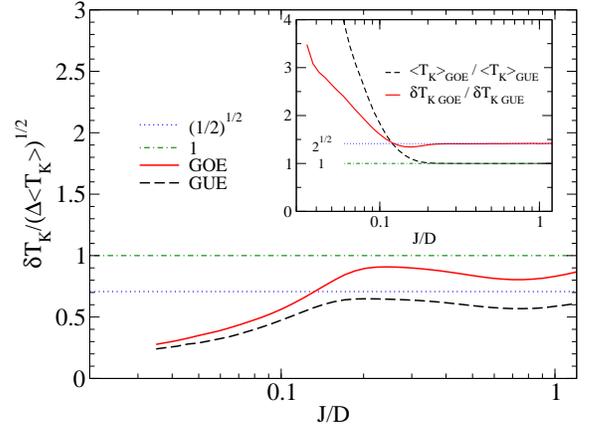}
\caption{(Color on-line) The ratio $\delta T_K/(\Delta \langle
T_K\rangle)^{1/2}$ as a function of the exchange coupling constant $J$
for the GOE and GUE ensembles. Inset: The ratio between GOE and GUE
values for $\langle T_K\rangle$ and $\delta T_K$. Analytical
considerations predict these ratios to yield the universal numbers $1$
and $\sqrt{2}$, respectively, over a wide range of exchange couplings
$J > J_-$. Deviations from this behavior are only seen for small
exchange couplings, when $J < J_-$. These can be attributed to
fluctuations of the two levels closest to the Fermi energy.}
\label{fig:dT_TK_RMT}
\end{figure}

Using the Gaussian approximation of Eq. (\ref{pgauss}), we can
establish a relation between the standard deviation and the average of
the Kondo temperature in the limit $T_K \gg \Delta$,
namely,\cite{kbu041,moriond,jetp}
\begin{equation}
\label{varrmt}
\delta T_K\mid_\beta \, \approx \sqrt{\frac{\langle T_K \rangle \,
\Delta}{\beta}},
\end{equation}  
where $\langle T_K \rangle \approx T_K^{(0)}$ is assumed. Since
$\langle T_K \rangle$ is ensemble independent in the $T_K \gg \Delta$
limit, one finds from Eq. (\ref{varrmt}) that $\delta T_{K\, {\rm
GOE}}/\delta T_{K\, {\rm GUE}} = 1/\sqrt{2}$. These results 
are compared to the numerical data in Fig. \ref{fig:dT_TK_RMT}. In
general, when time-reversal symmetry is present ($\beta=1$), level
repulsion is weaker and the tendency to localization
stronger. Consequently, the probability of having a vanishing wave
function at the magnetic impurity position is enhanced, as well as the
probability of large wave function splashes. This indicates that the
RMT distribution of $T_K$ should be wider in this case than for the
unitary class (broken time-reversal symmetry), in agreement with the
results of Ref. \onlinecite{jetp}. The inset of
Fig. \ref{fig:dT_TK_RMT} shows that, where applicable, the analytical
prediction for the {\it ratio} between the standard deviations of the
Kondo temperature for the orthogonal and unitary ensembles is in quite
good agreement with the numerical results. For both RMT ensembles the
distribution width scales with $\sqrt{\Delta}$ and therefore vanishes
in the infinite volume limit. In Sec. \ref{sec:disorderedmetals} we
showed that this is no longer true when spectral correlations beyond
the Thouless energy scale exist.

\begin{figure}[ht]
\vspace{1cm}
\includegraphics[width=7.5cm]{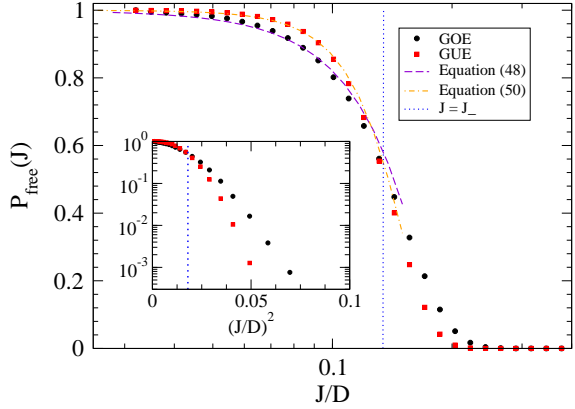}
\caption{(Color on-line) Probability of finding free magnetic
moments. The data points correspond to the fraction of events in the
RMT simulations where either $T_K=0$ or no solution to
Eq. (\ref{eq:sctk}) was found, and is plotted as a function of the
exchange coupling $J$. The dotted line indicates the point where
$J=J_-$ (clean case threshold). There is a good agreement between
simulations and the analytical predictions for $J \ll J_-$ based on
the two-level system model. The inset shows the tail of the
probability function at $J>J_-$.}
\label{noevents}
\end{figure}

There are realizations where no solution with $T_K>0$ can be found for
Eq. (\ref{eq:sctk}). They can be interpreted as the absence of
screening of the magnetic impurity dipole moment. For a clean system,
one can estimate the minimum exchange coupling $J_-$ where
Eq. (\ref{eq:sctk}) ceases to provide a finite value for $T_K$ (see
Appendix \ref{sec:clean}). Non-magnetic disorder strongly affects
$J_-$. For small exchange couplings, free moment events amount to a
large portion of all events. Their probability is plotted separately
in Fig. \ref{noevents}. It is remarkable that even for $J < J_-$,
where $\langle T_K \rangle \ll \Delta$, the distribution shows a clear
maximum at a finite value of $T_K$. The distribution (see
Ref. \onlinecite{jetp}) decays towards $T_K = 0$ even though there is
a large amount of free moments in this case (see
Fig. \ref{noevents}). This peculiar behavior can be qualitatively
understood by recalling the level repulsion of RMT ensembles and can
be quantitatively treated within a two-level model, as outlined in
Sec. \ref{sec:TLS} below.

\subsection{Two-Level Approximation ($T_K < \Delta$)}
\label{sec:TLS}
   
For $T_K < \Delta$, only the two states closest in energy to the Fermi
energy are appreciably coupled to the magnetic impurity. For small
exchange couplings $J < J_-$, this affects the whole distribution of
Kondo temperatures, while for larger exchange couplings it applies
only to the small-$T_K$ tail. In this limit, we can simplify the
analytical calculations by separating the contributions to the sum in
Eq. (\ref{eq:sctk}) into two groups. In the first we include only the
two nearest levels to the Fermi energy and fully take into account the
fluctuation of their relative spacing and their wave function
amplitudes. The remaining levels form the second group and are treated
assuming an equally spaced spectrum and $x_n = 1$. When the wave
functions at the Fermi level are distributed according to the GOE
Porter-Thomas distribution but the energy levels are kept fixed, the
average Kondo temperature is found to be
\begin{equation}
\langle T_K \rangle_{\rm GOE} = \frac{-{\rm Ei}(-1/2)}{2 e} \Delta
\exp \left( - \frac{D}{J} + \frac{D}{J_-} \right),
\label{tlsO}
\end{equation}
where ${\rm Ei}(x)$ denotes the exponential integral function,
\cite{gradstein} with ${\rm Ei}(-1/2) \approx -0.56$. For the GUE
Porter-Thomas distribution, we obtain instead
\begin{equation} 
\label{tlsU}
\langle T_K \rangle_{\rm GUE} = \frac{-2 {\rm Ei}(-1/2)}{e^2} \Delta
\exp \left( -\frac{2D}{J} + \frac{2D}{J_-} \right),
\end{equation} 
showing that the average Kondo temperature in the unitary ensemble
becomes exponentially smaller than in the orthogonal ensemble, which
is consistent with the numerical result presented in
Fig. \ref{fig:dT_TK_RMT}.

It is of particular interest to determine the probability $P_{\rm
free}(J)$ of having unscreened, free magnetic moments in the system
for a given bare exchange coupling. This function is equal to the
probability that the Kondo temperature is either zero or that there is
no solution to Eq. (\ref{eq:sctk}). In these cases, the correction to
the exchange coupling remains small for all temperatures and the Kondo
effect is quenched, in the sense that the magnetic moment remains
unscreened down to zero temperature. As discussed in Appendix
\ref{sec:clean}, in the clean limit the free moment probability is a
sharp step function of $J$, namely, $P_{\rm free}(J \le J_-)= 1$, and
$P_{\rm free}(J > J_-)=0$, with $J_-$ given by
Eq. (\ref{jminus}). When there is randomness, $J_-$ is a function of
the random wave function intensities $x_n$ and level spacings $s_n$ of
the closest levels to the Fermi energy. Upon averaging over their
distributions, we expect the step function to become a smoothly
decaying function. Thus, due to the finite probability that level
spacings at the Fermi energy exceed $\Delta$ or wave function
intensities are small, there is a finite probability of finding free
moments even at $J>J_-$, where the magnetic moment would be screened
without randomness. We have determined this probability distribution
numerically (see Fig. \ref{noevents}). For $J < J_-$, we can compare
it with the result obtained within a 2LS approximation. A first
attempt can be made by setting the wave function amplitudes to be
constant, $x_l =1$ and fixing all but the two closest levels to the
Fermi energy, which are allowed to fluctuate according to the Wigner
surmise.\cite{mehta} For the GOE we obtain
\begin{equation}
P_{\rm free} (J) = \exp \left[ - \frac{\pi}{(D/J - D/J_- + 2)^2}
\right],
\end{equation}
which is valid only for $J < J_+ = J_-/(1-J_-/D)$, since for larger
values of $J$ the fluctuations of the other levels can no longer be
neglected.
  
When the wave function amplitudes are constant, the probability to
have free magnetic moments is just proportional to the probability
that the level spacing at the Fermi energy is of the order of
$T_K^{(0)} (J)$. That is exponentially small for $J > J_-$. However,
in Fig. \ref{noevents} we see that the decay with $J$ is slower than
that, indicating that the fluctuations of wave functions are
crucial. Indeed, under realistic conditions, there is no appealing
reason to take $x_l=1$. When wave functions are allowed to fluctuate
according to the Porter-Thomas-distribution, we rather obtain (again
for the GOE)
\begin{eqnarray}
P_{\rm free}^{\rm GOE}(J) & = & \int_0^{\infty} dx\, \exp \left( -x -
\frac{\pi\, x^2}{4 u_J^2} \right) \nonumber \\ & = & u_J\, \exp \left(
\frac{u_J^2}{\pi} \right)\, {\rm Erfc} \left( \frac{u_J}{\sqrt{\pi}}
\right),
\end{eqnarray}
with 
\begin{equation}
u_J = \left| 1 + \frac{D}{2 J} -\frac{D}{2 J_-} \right|.
\end{equation}
Note that $P^{\rm GOE}_{\rm free}(0)=1$ and $P^{\rm GOE}_{\rm free}
(J_-) \approx 2/\pi$. For $J > J_-$, the fluctuations of the other
energy levels become important and the 2LS approximation can no longer
be used. Similarly, for the GUE we obtain
\begin{eqnarray}
P_{\rm free}^{\rm GUE}(J) & = & 1 - \frac{4}{\pi^2} \int_0^{\infty}
dx\, x^2\,  (1 + u_J x) \nonumber \\ & & \times \exp \left( - u_J x -
\frac{x^2}{\pi} \right).
\end{eqnarray}
These results are compared to the numerical calculations in
Fig. \ref{noevents} and good agreement is found for $J<J_-$. Thus, in
RMT, we can conclude that fluctuations of wave functions do enhance
the probability of finding free moments for $J > J_- $. 

Finally, there is remarkable feature of the distribution of the Kondo
temperature, as shown in Ref. \onlinecite{jetp}, especially noticeable
for small exchange couplings ($J/D < 0.1$): Although the number of
free moments is large in this regime, the distribution still has a
distinct maximum and turns to zero as $T_K \rightarrow 0$. We find
that this is due to the energy level repulsion of RMT. Evaluating the
distribution function in the 2LS approximation (i.e., for a fixed
energy level spacing but fluctuating wave functions), we arrive at
$P(T_K) \sim (\Delta/T_K^2) e^{-\Delta/2 T_K}$. Then, using the Wigner
surmise to average this expression over level spacings, we find that
the distribution decays as a power law,
\begin{equation} 
\label{rmtsmalltk}
P(T_K < \Delta) \sim \frac{a_\beta}{\beta + 2}
\frac{T_K^{\beta}}{\Delta^{\beta +1}},
\end{equation} 
with $a_1 = \pi/2$ and $a_2 = 32/\pi$, in fair qualitative agreement
with the numerical data presented in Ref. \onlinecite{jetp}.

\section{Dependence of $P(T_K)$ on the concentration of magnetic impurities}
\label{sec:concentration} 

In the absence of an external magnetic field, the symmetry class of
the underlying single-particle basis to the Kondo problem is
controlled by the concentration of magnetic impurities and the
relation between certain energy scales. The general idea is that, as
the number of magnetic impurities increases, statistical fluctuations
of the electron states in the sample cross over from the orthogonal
class (time-reversal symmetric) to the unitary class (broken
time-reversal symmetry). This is due to the fact that the spin
dynamics of the magnetic impurities can be slow compared to the time
scale of the conduction electrons, thereby breaking effectively the
time-reversal invariance on their time scale.\cite{hikami,bobkov}
However, as we will argue below, the dynamical regime in the sample,
and consequently the strength of the Kondo screening, also plays a
crucial role in determining on which side of this crossover the system
finds itself.\cite{meraikh}

We begin by recalling that for a mesoscopic sample in the weak
localization (WL) regime, i.e., at energies $E > E_c$ (short time
scales), the dimensionless parameter controlling the
orthogonal-unitary crossover due to magnetic impurities can be written
as \cite{hikami}
\begin{equation} 
X_s^{\rm WL} = \frac{1}{E_c\tau_s},
\end{equation} 
with the crossover centered at $X_s=1$. As the energy is lowered below
$E_c$ (long time scales), one enters the zero-dimensional, RMT regime,
where the average level spacing $\Delta$ takes over $E_c$ as the
relevant energy scale. In that regime, the crossover parameter is
given by
\begin{equation}
X_s^{\rm RMT} = \frac{1}{\Delta\tau_s} = {\cal G}\, X_s^{\rm WL}
\end{equation} 
instead. Thus, for a sample in the RMT regime, the crossover occurs
for spin scattering rates smaller by a factor $1/{\cal G}$ with
respect to the weak-localization regime, where ${\cal G}$ is the
dimensionless conductance. For the RMT regime, this crossover has been
recently studied for a Kondo quantum dot in an Aharonov-Bohm
ring.\cite{lw04}

At any finite temperature, the spin scattering rate $1/\tau_s$ is
renormalized by Kondo correlations: It is small at both $T\gg T_K$ and
$T\ll T_K$, having a maximum at around $T=T_K$, as has been observed
in weak-localization experiments.\cite{bergmann,mw00} This behavior is
analogous to that observed in the reentrance of gapless
superconductivity.\cite{maple} For a clean sample (see
Refs. \onlinecite{zarand,micklitz05}),
\begin{equation} 
\label{eq:tausoft}
\frac{1}{\tau_s(T)} = \left\{ \begin{array}{lr} \frac{\pi\, n_m\,
S(S+1)}{\nu}\, \ln^{-2}(T/T_K), & T > T_K, \\ \alpha\,
\frac{n_m}{\pi\,\nu}, & T \approx T_K, \\ \frac{n_m}{\pi\,\nu}\,
\left( \frac{T}{T_K} \right)^2 \frac{w^2 \pi^4}{16} c_{\rm FL}, & T
\ll T_K,
\end{array} \right.,
\end{equation}
where $n_m$ is the density of magnetic impurities, $S$ denotes their
spin, $\alpha$ is a numerical factor smaller than unit, and $w \approx
0.41$ is the Wilson number. In Refs. \onlinecite{zarand} and
\onlinecite{micklitz05}, it was found numerically that $\alpha \approx
0.2$. In Eq. (\ref{eq:tausoft}), $c_{\rm FL}$ is a factor that arises
from the relation between the inelastic scattering rate and the
temperature-dependent dephasing rate.\cite{micklitz05} In 2D, $c_{\rm
FL} \approx 0.946$.

A few elucidating remarks about Eq. (\ref{eq:tausoft}) are in
order. First, we define the spin scattering rate as in
Ref. \onlinecite{hikami}, which is smaller than the dephasing rate
defined in Ref. \onlinecite{micklitz05} by a factor 1/2. Secondly,
note the difference between our definition of $T_K$ and that used in
Ref. \onlinecite{micklitz05}: Ours corresponds to their {\it
perturbative} $T_K$. Thirdly, the numerical prefactor of the
$(T/T_K)^2$ term in Eq. (\ref{eq:tausoft}) was obtained within Fermi
liquid theory,\cite{hewson} where the exact result differs from that
obtained in the standard Sommerfeld low-temperature expansion by a
factor $3$. In the following, we will approximate the numerical
prefactor multiplying the $(T/T_K)^2$ term by unit, since $w^2 \pi^4
c_{\rm FL}/16 = 0.968$.

The low-temperature limit of the spin scattering rate shows the
expected Fermi-liquid scaling based on Nozieres' theory for the Kondo
problem, which is valid at $T\ll T_K$.\cite{nozieres,affleck} At
temperatures exceeding $T_K$, Eq. (\ref{eq:tausoft}) is consistent
with the perturbative poor man's
scaling.\cite{anderson,abrikosov,suhl,zittartz,maple} However, we note
that recent experiments have shown the scattering rate to be
approximately linear with temperature for a wide interval below
$T_K$.\cite{sbs03} This behavior has been explained theoretically by
Zarand and coworkers using a numerical renormalization group
calculation of the frequency-dependent inelastic scattering
rate,\cite{zarand} where they also obtained $\alpha \approx 0.2$.

In principle, one could expect that the maximum value $1/\tau_s$ can
reach should be given by the unitary limit of the scattering cross
section. However, in reality, the maximum value is found to be smaller
than that by the factor $\alpha$ shown in Eq. (\ref{eq:tausoft}). In
2D,\cite{zarand}
\begin{equation} 
\left.\frac{1}{\tau_s}\right|_{\rm max} = \alpha\, \frac{n_m}{\pi \nu},
\end{equation} 
Thus, the maximal value the crossover parameter can reach for a
two-dimensional sample in the weak localization regime is
\begin{equation} 
\left. X^{\rm WL}_s \right|_{\rm max} = \alpha \frac{N_m}{\pi {\cal
G}},
\end{equation}
where $N_m = n_m L^2$ is the number of magnetic impurities for a
sample with linear size $L$. When there are only a few magnetic
impurities, $N_m < {\cal G}$, the crossover parameter is small and the
sample is in the orthogonal regime. Increasing the concentration of
magnetic impurities increases the parameter $X_s$ and eventually leads
to the unitary regime. As found in Sec. \ref{sec:rmt}, this is
accompanied by a decrease in the width of the distribution of Kondo
temperatures.

It has recently been pointed out that the correlations between wave
functions at different energies are enhanced in the GOE-GUE crossover
regime.\cite{adam} Thus, since the width of the distribution of the
Kondo temperature is enhanced by wave function correlations, one can
expect a widening of the Kondo distribution in the GOE-GUE crossover
regime due to the presence of a weak magnetic field or a small amount
of spin scattering.\cite{kravtsov} The wave function correlation
function is then given by
\begin{equation}
\Omega^2 \left. \left\langle \mid \psi_n( {\bf r}) \mid^2 \mid \psi_m(
{\bf r}) \mid^2 \right\rangle \right|_{\omega = E_n - E_m} = \frac{2
\lambda^2 \Delta^2}{4 \lambda^4 \Delta^2 + \pi^2 \omega^2},
\end{equation}
Here, $\lambda$ is a dimensionless parameter related to the magnitude
of the time-reversal breaking perturbation. In the presence of an
external magnetic field $B$, $\lambda^2 = 1/\tau_B\Delta$, where
$1/\tau_B$ is the magnetic phase-shift rate. For a diffusive quantum
dot, this rate is given by $1/\tau_B = \gamma e^2 D_e B^2 L^2/h^2 $,
where $\gamma$ is a geometrical factor of order unity. In the presence
of magnetic impurities, the GOE-GUE crossover is governed by the spin
scattering rate and $\lambda^2 = 1/\tau_s\Delta$. Remarkably, the
mixing of GOE eigenfunctions brought by time-reversal symmetry
breaking induces wave function correlations over a macroscopic energy
scale equal to either $1/\tau_B$ or $1/\tau_s$. However, the amplitude
of these correlations is proportional to $ \tau_s\Delta$ and therefore
they vanish in the thermodynamic limit.\cite{adam} In order to see
that, we use Eq. (\ref{eq:TK_transcd}) to explicitly evaluate the
width of the distribution of the Kondo temperature related to the
crossover wave function correlations. We obtain
\begin{equation}
\left\langle \delta T_K^2 \right\rangle = 2 \frac{\Delta}{\tau_s}
\int_0^{1/(1+\pi T_K^{(0)} \tau_s)} \frac{d t}{t} \ln \frac{1+t}{1-t}.
\end{equation}
Thus, for $T_K^{(0)} \gg 1/\tau_s$, these GOE-GUE correlations only
contribute to a width $\delta T_K \approx T_K^{(0)}
\sqrt{\tau_s\Delta/\pi}$ that vanishes in the thermodynamic limit when
$\Delta \rightarrow 0$, but is larger than the value obtained in the
pure RMT ensembles by a factor $\sqrt{T_K^{(0)} \tau_s}$

As the magnetic impurity concentration increases further, the
superexchange interaction between magnetic impurities begins to
compete with the Kondo screening.\cite{RKKY} The superexchange
interaction coupling $J_{\rm RKKY}$ fluctuates and its average and
spreading depend on the amount of disorder, impurity concentration,
and details of the Fermi surface.  $J_{\rm RKKY}$ is zero on
average,\cite{degennes} but fluctuates according to a wide log-normal
distribution.\cite{lerner} Its typical value is of the same order as
that of a clean sample,\cite{zyuzin} namely, $\sqrt{\langle J^2_{\rm
RKKY}\rangle}= (n_m/\nu) (J/D)^2 \cos (2 k_F R)$. Even when the
typical superexchange coupling constant is smaller than $T_K$, there
is a small chance that clusters of localized spins form. When two
localized spins couple ferromagnetically to form a triplet state, they
contribute to the dephasing rate of itinerant electrons. Such a
contribution to the dephasing rate scales with
$n_m^2$.\cite{frossati,micklitz05} When $J_{\rm RKKY}$ exceeds the
Kondo temperature $T_K$ of a single magnetic impurity, the spins of
the magnetic impurities are quenched. They form a classical random
spin array whose spin scattering rate is smaller than that for free
quantum spins by the factor $1/3$, and scale linearly with
$n_m$.\cite{spinglass} At intermediate concentrations of magnetic
impurities Griffiths-McCoy singularities can appear and induce an
interesting though more complex behavior (see
Ref. \onlinecite{castroneto} for a review).

\section{Dephasing due to Free Magnetic Moments in Disordered Metals}
\label{sec:detection}

We now discuss the relevance of the distribution of the Kondo
temperature $T_K$ to the quantum corrections of the conductance of
mesoscopic wires.

As discussed in Sec. \ref{sec:concentration} [see
Eq. (\ref{eq:tausoft}), in particular], the temperature dependence of
the spin scattering rate has a maximum at $T_K$, resulting in a
plateau of the dephasing time.\cite{bergmann,mw00} However, we have
found in Sec. \ref{sec:numerics} that disorder can result in a finite
probability of having magnetic impurities with vanishingly small Kondo
temperatures. Thus, the question arises whether the latter effect can
yield a finite contribution to the dephasing rate at temperatures
below the average Kondo temperature.  It has recently been argued that
the dephasing rate due to a magnetic impurity can be related to the
inelastic cross section of a magnetic impurity at the Fermi
energy,\cite{zarand,micklitz05}
\begin{equation} \label{taus} 
\frac{1}{\tau_{s}} = n_m v_F  \sigma_{\rm inel},
\end{equation}
where $v_F$ is the Fermi velocity, and the inelastic cross section can
be obtained from the difference between the total and the elastic
cross section,
\begin{equation}
\sigma_{\rm inel} (E_F)  = \sigma_{\rm total} - \sigma_{\rm el}.
\end{equation}
Using the optical theorem, both the total and the elastic cross
sections can be related to matrix elements of the $T$-matrix of the
magnetic impurity.\cite{zarand} Based on the fact that in the
derivation of the weak-localization correction one averages over the
disorder potential and thereby recovers translational invariance, the
authors of Refs. \onlinecite{zarand} and \onlinecite{micklitz05}
calculated the spin scattering rate of Eq. (\ref{taus}) for a plane
waves basis. However, one should calculate the spin scattering rate as
averaged over all momentum directions in order to correctly determine
how the dephasing rate caused by magnetic impurities is affected by
the random environment. Following this approach, we find that the spin
scattering rate of a state of energy $E$ should be given by\cite{obs3}
\begin{eqnarray}
\label{1tausplain}
1/\tau_s (E , T) = \frac{n_m}{\nu} \sum_p \delta (E_p - E) \Big[
{\rm Im}\, \langle p | T | p \rangle \nonumber \\ - \frac{\pi}{\Omega}
\sum_{p^\prime} \delta(E_p -E_{p^\prime})\, |\langle p | T | p^\prime
\rangle |^2 \Big].
\end{eqnarray} 
Using the relation between the $T$-matrix element and the propagator
of a localized $d$-level in the Anderson model,\cite{zarand}
$G_d(E_n,T)$, we obtain
\begin{equation}
\langle p |T | p^\prime \rangle = - \Omega\, \psi_{p^\prime}(0)\,
\psi^\ast_p(0)\, t_0^2\, G_d.
\end{equation} 
Inserting this relation into Eq. (\ref{1tausplain}), we find that the
spin scattering rate is given by
\begin{eqnarray}
\label{tausdisordered}
1/\tau_s (E , T) & = & - \, n_m\, \Omega \frac{\rho(E,0)}{\nu} \left(
t_0^2\, {\rm Im}\, G_d + \pi \nu\, t_0^4\, G_d^2 \right) \nonumber \\
& = & \frac{\rho(E,0)}{\nu} 1/\tau_s^{(0)} (E,T),
\end{eqnarray}
where $1/\tau_s^{(0)}$ is the inelastic spin scattering rate in a
clean system. $1/\tau_s^{(0)}(E,T)$ has a universal dependence on
$T/T_K$, as has been shown in Refs. \onlinecite{zarand} and
\onlinecite{micklitz05}. Thus, Eq. (\ref{tausdisordered}) leads us to
conclude that the spin scattering rate not only depends on the ratio
$T/T_K$ but also explicitly depends on the local density of
states.  The same  conclusion has been 
 recently reached independently in Ref. (\onlinecite{micklitz06}). But both $T_K$ and $\rho(E,0)$ are randomly distributed,
taking different values for each magnetic impurity, and so will
$1/\tau_s(T)$. Note that the distribution of Kondo temperatures alone
is not sufficient for determining the average spin scattering rate. We
have seen in Sec. \ref{sec:disorderedmetals} that in a two-dimensional
metal, the wave functions correlations at different energies are of
order $1/g$. The probability of having simultaneously a small Kondo
temperature and a small local density of states at the Fermi energy
leading to an anomalous value for the spin scattering rate is thus
expected to be of order $1/g$ as well.

\begin{figure}[h]
\includegraphics[width=7.3cm]{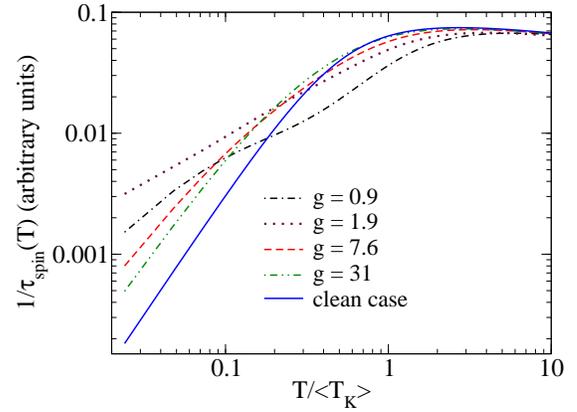}
\caption{The spin relaxation rate as a function of temperature for
$J/D = 0.24$. The broken lines correspond to ensemble averages of
Eq. (63) over fluctuating Kondo temperatures and the LDOS at the Fermi
energy (data from the $20 \times 20$ lattice simulations was
used). The solid line is obtained from Eq. (64) setting $T_K = \langle
T_K \rangle$ (clean case).}
\label{fig:spinrelaz}
\end{figure}

We can derive the temperature dependence of the spin scattering rate
in a sample with a finite number of magnetic impurities by averaging
the dephasing rate of a single magnetic impurity over an ensemble of
magnetic impurities. To facilitate the calculations, we approximate
the universal function for the dephasing rate by the following
function,
\begin{eqnarray}
\label{eq:tauapprox}
\frac{1}{\tau_s^{(0)}} & = & \frac{\pi\, n_m\, S(S+1)}{\nu} \left\{
\ln^2 \left( \frac{T}{T_K} \right) \right. \nonumber \\ & & \left. +
\pi^2 S (S+1) \left[ \left( \frac{T_K}{T} \right)^2 + \frac{1}{\alpha}
- 1 \right] \right\}^{-1},
\end{eqnarray}
with $\alpha = 0.2$, as found numerically in
Ref. \onlinecite{zarand}. Note that temperature scales with $T_K$
only. Equation (\ref{eq:tauapprox}) coincides with
Eq. (\ref{eq:tausoft}) in both low- and high-temperature limits. In
Fig. \ref{fig:spinrelaz} we show the result of using
Eq. (\ref{eq:tauapprox}) in Eq. (\ref{tausdisordered}) and ensemble
averaging the spin relaxation rate over $T_K$ and $\rho(E_F,0)$
obtained numerically for the $20 \times 20$ lattice model (see
Sec. \ref{sec:numerics}). 

One finds a clear departure from the clean case prediction at low
temperatures,\cite{micklitz05} which amounts to an enhancement of the
spin scattering due to the presence of nonmagnetic disorder.
Moreover, the maximum value of the spin scattering rate is found to be
reduced in proportion to the disorder strength and its position
shifted to larger temperatures. A very similar behavior has been
recently observed in weak-localization measurements performed by the
Grenoble group on quasi-one-dimensional Ag wires doped with 2 to 20
ppm Fe impurities,\cite{mallet06} as well as by the Michigan
group,\cite{birge06} where similar samples where implanted with 2 to
10 ppm of Fe impurities. One should note that these samples have very
high diffusion constants, of order $D_e = 300-400$ cm$^2$/s, so that
the parameter $g$ in these samples is large, $g > 100$. For the Fermi
energy in Ag, $E_F = 5.5$ eV, and the Kondo temperature of $T_K= 4$ K,
the ratio between band width and exchange coupling is found to be of
the order of $J/D = 0.1$. Thus, according to Eq. (\ref{2dwidth}),
$\delta T_K/ \langle T_K \rangle \ll 1$ is fulfilled and the width of
the distribution of Kondo temperatures should be negligible in 2D
samples. However, one has to be careful when considering the relevance
of our results to the weak-localization experiments of
Refs. \onlinecite{mallet06} and \onlinecite{birge06}. For instance, we
note that the low-temperature dephasing rate is dominated by the
non-Gaussian low-$T_K$ tail of the distribution of the
Kondo-temperature, which we have found to be present even for weak
disorder, $g \gg 10$. Moreover, the samples used in
Refs. \onlinecite{mallet06} and \onlinecite{birge06} were
quasi-one-dimensional, in which case $\delta T_K$ is known to be
further enhanced (see Sec. \ref{sec:disorderedmetals}).

In order to connect theory with experiments, some more work needs to
be done. On the theoretical side, a systematic study of how the
distribution of the Kondo temperature scales with the width and length
of the sample is necessary. This is the subject of ongoing work and
will be published elsewhere.\cite{future} On the experimental side,
the relevance of the distribution of the Kondo temperature to the
low-temperature anomaly of the dephasing rate could be established by
utilizing samples with lower diffusion constants, such as the
low-mobility samples of AuPd, examined in Ref. \onlinecite{natelson}
(see also Ref. \onlinecite{lin}). These have diffusion constants as
low as $D_e= 13.4$ cm$^2$/s, corresponding to $g \approx17$. The
presence of free magnetic moments would be facilitated in
semiconductors such as Si and GaAs where the effective mass of
electrons is smaller than in metals by a factor 30. Moreover, 2D
electron gases with $a k_F \approx 1$ can be produced with those
materials. In low-mobility GaAs wires, Anderson localization has been
observed even in wires with $g \approx 30$.\cite{gershenson} Thus, all
mesoscopic energy scales can become relevant when the system reaches
temperatures of the order of 1 K. However, so far no magnetic
impurities with detectable Kondo temperatures have been found in
semiconductors, since the low density of states suppresses the Kondo
temperature exponentially.

\section{Conclusions} 
\label{sec:conclusions}

The screening of magnetic moments in metals, the Kondo effect, is
found to be quenched with a finite probability in the presence of
nonmagnetic disorder. For weak disorder, $g \gg 1$, the effect is
shown analytically and numerically to be due to wave function
correlations. Albeit weak in the metallic regime, these correlations
have a sizeable effect on the Kondo temperature. That is because the
Kondo temperature involves a summation over all eigenstates with a
finite wave function amplitude at the location of the magnetic
impurity. As disorder increases and the sample goes from the diffusive
to the localized regime, an increase in the correlations of the LDOS
at different energies induces even stronger fluctuations in the Kondo
temperature. Thus, the distribution of the Kondo temperature $P(T_K)$
retains a finite width in the limit of vanishing level spacing
$\Delta$. When time-reversal symmetry is broken either by applying a
magnetic field or by increasing the concentration of magnetic
impurities, $P(T_K)$ becomes narrower. The probability that a magnetic
moment remains free down to the lowest temperatures is found to
increase with disorder strength. This result is consistent with direct
measurements of Kondo effect in Cu(Fe) thin films and Au(Fe)
wires.\cite{giordano}

We have also shown that magnetic impurities with a small Kondo
temperature are shown to modify the temperature dependence of the
dephasing rate at low temperatures, $T \ll \langle T_K \rangle$, as
measured in weak-localization transport experiments.

Equations (\ref{eq:lnTK}) and (\ref{eq:lnTKQ1D}) allow us to conclude
that the disorder can substantially affect the Kondo temperature. This
could be contrasted with the effect of disorder on another energy
scale derived from a many-body state, namely, the critical temperature
of a superconductor, $T_c$. Anderson's theorem states that $T_c$ is
not affected by nonmagnetic disorder.\cite{rickayzen} Our results
demonstrate that there is no analog to this theorem for the Kondo
temperature of magnetic impurities embedded in low-dimensional
disordered systems since the variance of $T_K$ depends markedly on the
disorder strength. This conclusion, however, is at odds with the
proposal of Ref. \onlinecite{chakravarty00}, where it was argued that
in the low-temperarature, strong coupling limit, the Kondo effect
is not affected by disorder at all. We note that this result was
obtained using a path integral representation and a semiclassical
expansion (the large-$N$ limit is taken, where $N=2s+1$ is the number
of spin components). The disagreement between
Ref. \onlinecite{chakravarty00} and our results may not be real,
however, since their calculation is limited to $T \ll T_K$ and cannot
rule out the existence of magnetic impurities with exceedingly small
$T_K$.

It remains to be studied how the distribution of $T_K$ scales with
system size and how it changes (if at all) in the
quasi-one-dimensional limit. The numerical techniques employed in this
work did not permit the investigation of sufficiently large
lattices. It also remains to be checked whether the large fluctuations
of $T_K$ seen at strong disorder survive a non-perturbative approach
such as the numerical renormalization group. Finally, we note that we
have not attempted to explore the consequences of having a free
magnetic moments to the nature of the electron liquid. This issue has
been recently addressed by other authors in the context of
heavy-fermion materials.\cite{miranda,castroneto}

\section*{Acknowledgments} 

The authors gratefully acknowledge useful discussions with Harold
Baranger, Carlo Beenakker, Antonio Castro Neto, Claudio Chamon,
Vladimir Dobrosavljevic,  Peter Fulde, Ribhu Kaul, Vladimir Kravtsov,
Caio Lewenkopf, Alexander Lichtenstein, Ganpathy Murthy, Mikhail
Raikh, Achim Rosch, Denis Ullmo, Chandra Varma, Isa Zarekeshev, and Andrey
Zhuravlev. The authors thank the hospitality of the Condensed Matter
Theory Group at Boston University, where this work was initiated, and
the Aspen Center for Physics, where the manuscript was
finalized. S.K. also acknowledges the hospitality of the Max-Planck
Institute for Physics of Complex Systems and the Department of Physics
at the University of Central Florida. E.R.M. thanks the hospitality of
the Institute for Theoretical Physics at the University of
Hamburg. This research was supported by the German Research Council
(DFG) under SFB 508, A9, and SFB 668, B2.

\appendix
\section{Derivation of Eq. (\ref{eq:tk})}
\label{sec:appendixA}

We begin with the Anderson model, namely, a localized $d$-level
coupled to a conduction band,\cite{anderson}
\begin{eqnarray}
H &=& \sum_{n,\sigma} E_n \hat{n}_{n \sigma} + \varepsilon_d
 \sum_{\sigma} \hat{n}_{d \sigma} + U \hat{n}_{d +} \hat{n}_{d -}
 \nonumber \\ &+& \sum_{n,\sigma} \left( t_{n d} c^+_{n \sigma} c_{d
 \sigma } + t_{d n} c^+_{d \sigma} c_{n \sigma } \right),
\end{eqnarray}
where $E_n$ are the eigenenergies of the conduction band electrons,
$\varepsilon_d$ is the energy of a singly occupied $d$-level, and
$\varepsilon_d + U$ is the energy of a doubly occupied
$d$-level. $t_{d n} = t_{n d}^*$ are the hybridization matrix elements
between the $d$-level and the conduction band state
$|n\rangle$. Projecting out of the Hilbert space all state where the
$d$-levels are doubly occupied one obtains the Kondo Hamiltonian
\begin{equation}
H_J = J\, {\bf s} \cdot {\bf S},
\end{equation}
where the matrix elements of the exchange interaction in the basis of
the eigenstates of the conduction electrons are given by
\begin{equation}
J_{kl} = \frac{8\, t_{k d}\, t_{d l}}{U}.
\end{equation}
These matrix elements are positive and thus the interaction is
antiferromagnetic.
The hopping matrix element connecting the localized $d$-state
$\phi_d({\bf r})$ to the conduction band state $\psi_n({\bf r})$ is
given by
\begin{equation}
t_{d n} = \int d^dr\, \phi_d^\ast({\bf r})\, \hat{V({\bf r})}\,
\psi_n({\bf r}),
\end{equation}
where $\hat{V}$ is the potential energy of the tunneling barrier. For
an impurity state strongly localized at ${\bf r}=0$ we obtain
\begin{equation}
J_{kl} \approx J\, \phi^\ast_d(0)\, \psi_n(0),
\end{equation}
where $J= 8 (1/m^\ast a_0^2)^2 U $, $m^\ast$ is the band mass, and
$a_0$ is the radius of the localized state at the magnetic atom. We
define $T_K$ by the divergence of second-order perturbation theory. To
second order in $J$, there are two processes to be considered: (i) The
scattering due to the exchange coupling $J$ of an electron from state
$|n\rangle$ to a state $|l\rangle$ close to the Fermi energy via an
intermediate state $|m\rangle$. This process is proportional to the
probability that state $|m\rangle$ is not occupied, $1-f(E_m)$, where
$f(E)$ is the Fermi distribution. (ii) The reverse process, in which a
hole is scattered from the state $|l\rangle$ to the state $|n \rangle$
via the occupied of a state $|m\rangle$ with probability proportional
to the occupation factor $f(E_m)$. Thus, we find that the exchange
coupling is renormalized to
\begin{equation}
\label{eq:poorman}
\tilde{J}_{nl} = J_{nl} \left[ 1 + \frac{J}{2 N} \sum_m \frac{
\Omega | \psi_m(0) |^2 }{E_m - E_F}
 \tanh \left( \frac{E_m - E_F}{2 T} \right) \right].
\end{equation}
For positive exchange coupling, $J>0$, perturbation theory diverges as
the temperature is lowered. Defining the Kondo temperature as the
temperature where the second-order correction to the exchange coupling
becomes equal to the bare coupling, we arrive at Eq. (\ref{eq:tk}).

An equivalent expression can be derived from the renormalization group
equation, which in the two-loop approximation is given by \cite{zu96}
\begin{eqnarray}
\label{eq:twoloop}
\frac{d J}{d t} & = & J^2\frac{\rho(E_F + \Lambda) + \rho(E_F -
\Lambda)}{2 D \nu} \nonumber \\ & & -\frac{1}{2} \frac{J^3}{D^2}
\frac{\rho(E_F-\Lambda)}{\nu} \int_0^{\Lambda}\frac{dE}{\Lambda}
\frac{\rho(E_F + E)/\nu}{(1+ E/\Lambda)^2} \nonumber \\ & &
-\frac{1}{2} \frac{J^3}{D^2} \frac{\rho(E_F+\Lambda)}{\nu}
\int_0^{\Lambda}\frac{dE}{\Lambda} \frac{\rho(E_F - E)/\nu}{(1+
E/\Lambda)^2}, \nonumber \\
\end{eqnarray}
where $t = \ln D/(2 \Lambda)$. Defining $T_K$ as the value of
$\Lambda$ at which $J$ flows to the band width $D$ and solving
Eq. (\ref{eq:twoloop}) for $T_K$, we obtain for a clean system
\begin{equation}
T_{K\, {\rm 2-loop}}^{(0)} = \frac{e}{2^{3/2}} \sqrt{\frac{J}{D}} \exp
(-D/J).
\end{equation}
However, if we keep only the first term on the right-hand side of
Eq. (\ref{eq:twoloop}) - the one-loop term, we obtain a
self-consistency equation for the Kondo temperature that reads
\begin{eqnarray}
\label{eq:RG1loop}
\ln \left[ \frac{T_K}{T_{K}^{(0)}} \right] & = & \frac{1}{2} \int_{2
T_K/D < \mid t \mid < 1} \frac{d t }{t} \left[ \frac{\delta \rho(E_F+
D t/2)}{\nu} \right].
\end{eqnarray}
For $T_K \gg \Delta$, Eq. (\ref{eq:RG1loop}) coincides with the
self-consistency equation obtained using perturbation theory, namely,
Eq. (\ref{eq:tk}).

\section{Clean case}
\label{sec:clean}

We briefly describe the behavior of the Kondo temperature in the
clean, bulk limit. For that purpose, we assume a spectrum of $N$
equally spaced levels, band width $D=N\Delta$, and spatially uniform
wave function intensities (plane waves). When the Fermi energy is in
the middle of the band and $N$ is even, all levels are either doubly
occupied or empty at $T=0$ and we have $s_n = n - N/2 -1/2$, $n =
1,\ldots,N$. For $N \gg 1$ and $T_K \gg \Delta$, we then find for
Eq. (\ref{eq:tk}) the well-known solution
\begin{equation}
\label{eq:tkusual}
T_K^{(0)} \approx 0.57\, D \exp ( - D/J), 
\end{equation}
which agrees to lowest order in $J/D$ with results from more accurate
methods such as the numerical renormalization group. In fact, for the
latter, the next leading order correction in the exponent has been
found to be $- 0.5 \ln (D/J)$,\cite{wilson} indicating that our
treatment is valid for $J < D$ up to preexponential corrections. For
small $J$, however, $T_K$ approaches $\Delta$ and turns abruptly to
zero at
\begin{equation}
\label{jminus}
 J_{-} = \frac{D}{\ln (2 N) + C}
\end{equation} 
in a nonanalytical fashion, namely,
\begin{equation}
\label{tksmall}
T_K^{(0)}(J\rightarrow J_-) = -\frac{\Delta}{2} \frac{1}{\ln
\left[(D/J_- - D/J)/4\right]}
\end{equation}
(here $C \approx 0.58$ is the Euler number). For $J < J_-$,
Eq. (\ref{eq:tk}) has no solution in the clean limit. That means that
the lowest order corrections to the bare exchange coupling remain weak
down to zero temperature and marginal terms have to be included in
order to determine $T_K$.\cite{wilson} Nevertheless, $T_K \ll \Delta$
in this regime and we can conclude that there will be unscreened
magnetic impurities (free moments) down to the lowest accessible
temperatures in bulk clean metals when $J < J_-$.




\begin{thebibliography}{11}
\bibitem{obs2} We did not use the eigenfunctions resulting directly
from the diagonalization of the GOE and GUE matrices to compute the
Kondo temperature for the following reason. Even though these
eigenfunctions do obey the Porter-Thomas distributions individually,
they are constrained by the normalization condition $\sum_{n=1}^N x_n
= N$. That alone makes $\delta T_K \rightarrow 0$ as $J$ approaches
$D$ and $T_K$ itself becomes of order $D$. While this is correct for a
single-band lattice model, it is not adequate for describing a
multi-band continuous system.
\bibitem{obs3} Note that the spin scattering rate in
Eq. (\ref{1tausplain}) is $1/2$ of the dephasing rate as defined in
Ref. \onlinecite{micklitz05}.
\end{thebibliography}
\end{document}